\RequirePackage{amsmath} 
\documentclass{article}

\usepackage[preprint]{neurips_2023}

\usepackage[utf8]{inputenc} 
\usepackage[T1]{fontenc}    
\usepackage{hyperref}       
\usepackage{url}            
\usepackage{booktabs}       
\usepackage{amsfonts}       
\usepackage{nicefrac}       
\usepackage{microtype}      
\usepackage{xcolor}         

\usepackage[pdftex]{graphicx}
\usepackage{tabularx}
\usepackage{subcaption}
\usepackage{multirow}
\usepackage[normalem]{ulem}
\usepackage{float}
\usepackage[inline]{enumitem}
\usepackage[capitalise]{cleveref}
\usepackage{caption}
\usepackage{subcaption}

\usepackage{pgfplots}
\usepackage{tikz}
\usetikzlibrary{arrows,automata,positioning}

\usepackage{pifont}
\newcommand{\cmark}{\ding{51}}%
\newcommand{\xmark}{\ding{55}}%

\newcommand{\gcell}{\cellcolor{green!25}}
\newcommand{\rcell}{\cellcolor{red!25}}

\definecolor{lightblue}{rgb}{0.88, 0.96, 1}
\newcommand{\colorours}{\rowcolor{lightblue}}

\usepackage[normalem]{ulem}
\usepackage{colortbl}

\newif\ifdraft
\drafttrue 

\ifdraft
    
    \newcommand{\whr}[1]{\textcolor{magenta}{\sout{#1}}}
    \newcommand{\whc}[1]{\textcolor{magenta}{[WN: #1]}}
    \newcommand{\mlc}[1]{\textcolor{magenta}{[ML: #1}}
    \newcommand{\bsc}[1]{\textcolor{blue}{[BS: #1]}}
    \newcommand{\bck}[1]{\textcolor{green}{[BK: #1]}}
    \definecolor{av_comment}{RGB}{255, 128, 0}

\else
    
    \newcommand{\whr}[1]{}
    \newcommand{\whc}[1]{}
    \newcommand{\mlc}[1]{}
    \newcommand{\bsc}[1]{}
    \newcommand{\bck}[1]{}
\fi

\newcommand{\vb}{Voicebox}

\newcommand{\pt}{p_t(x)}
\newcommand{\cpt}{p_t(x \mid x_1)}
\newcommand{\cvf}{u_t(x \mid x_1)}

\newcommand{\xctx}{x_{ctx}}
\newcommand{\xmis}{x_{mis}}
\newcommand{\lctx}{l_{ctx}}
\newcommand{\lmis}{l_{mis}}

\newcommand{\lhmis}{\hat{l}_{mis}}

\newcommand{\xhmis}{\hat{x}_{mis}}

\newcommand{\boldparagraph}[1]{\noindent{\bf #1:}}
\newcommand{\R}[2][]{\mathbb{R}_{#1}^{#2}}

\title{\vb{}: Text-Guided Multilingual \\Universal Speech Generation at Scale}

\author{
\centering 
\centerline{
Matthew Le\thanks{Equal contribution. Corresponding authors: \texttt{\{mattle,wnhsu\}@meta.com}}\quad
Apoorv Vyas$^{*}$\quad
Bowen Shi$^{*}$\quad
Brian Karrer$^{*}$\quad
Leda Sari\quad
Rashel Moritz\quad
}\vspace{2mm}
\and
\centerline{\textbf{
Mary Williamson\quad
Vimal Manohar\quad
Yossi Adi\thanks{FAIR \& Hebrew University of Jerusalem.}\quad
Jay Mahadeokar\quad
Wei-Ning Hsu$^{*}$\quad
}}
\vspace{4mm}
\and
{\normalfont Fundamental AI Research (FAIR), Meta}
}

\begin{document}
\maketitle

\begin{abstract}
Large-scale generative models such as GPT and DALL-E have revolutionized natural language processing and computer vision research. These models not only generate high fidelity text or image outputs, but are also generalists which can solve tasks not explicitly taught. In contrast, speech generative models are still primitive in terms of scale and task generalization.
In this paper, we present \vb{}, the most versatile text-guided generative model for speech at scale. 
\vb{} is a non-autoregressive flow-matching model trained to infill speech, given audio context and text, trained on over 50K hours of speech that are neither filtered nor enhanced.
Similar to GPT, \vb{} can perform many different tasks through in-context learning, but is more flexible as it can also condition on future context. \vb{} can be used for mono or cross-lingual zero-shot text-to-speech synthesis, noise removal, content editing, style conversion, and diverse sample generation. 
In particular, \vb{} outperforms the state-of-the-art zero-shot TTS model VALL-E on both intelligibility (5.9\% vs 1.9\% word error rates) and audio similarity (0.580 vs 0.681) while being up to 20 times faster. 
Audio samples can be found in \url{https://voicebox.metademolab.com}.
\end{abstract}

\section{Introduction}

\begin{figure}[h]
    \centering
    \includegraphics[width=.88\linewidth, trim=0 150 260 0, clip]{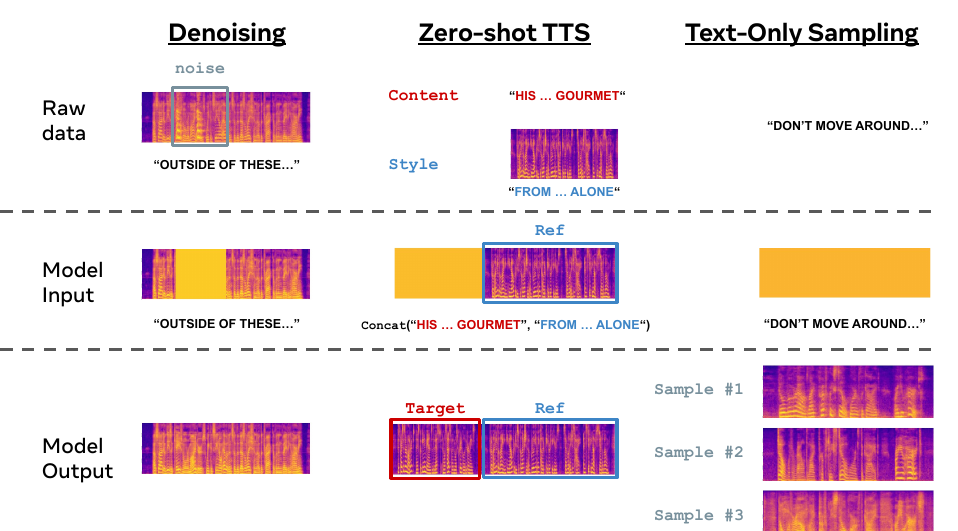}
    \caption{\vb{} task generalization through in-context learning.}
    \label{fig:app}
\end{figure}

Recent advances in large-scale generative models~\citep{Brown2020LanguageMA, Nichol2021GLIDETP, Ramesh2021ZeroShotTG} have led to a major paradigm shift towards building general-purpose models, which can perform many new tasks not explicitly trained on. These generative models learn to predict the missing data given the context.
Post training, we can directly input a question, optionally with a few contextual question-answer examples, instead of fine-tuning with labeled data. 
To give a concrete example, the model can answer a question like, ``What is the capital of Japan?'' with an example of such a relationship in the context: ``The capital of Germany is Berlin. The capital of Japan is''.

While the training objective appears simple, it subsumes many tasks as one can convert them into some form of context. For the model to perform well at every task, it implies that the estimation of $p(\text{missing data} \mid \text{context})$ needs to be accurate for every context. Hence, scale and diversity are the most crucial factors for building general-purpose models~\citep{Hoffmann2022TrainingCL,Aghajanyan2023ScalingLF}, as we see the state-of-the-art (SOTA) language and vision-language models are trained on web-scale data with billions to hundreds of billions of parameters.

Despite the success of large-scale generative models in other areas, most speech models are still trained on datasets at the scale of tens to hundreds of hours~\citep{Ren2020FastSpeech2F,Kim2020GlowTTSAG,Kim2021ConditionalVA,Popov2021GradTTSAD,Huang2022FastDiffAF,Tan2022NaturalSpeechET,Casanova2021YourTTSTZ}. Previous works consider highly curated datasets such as VCTK~\citep{Yamagishi2019CSTRVC}, which contains only clean audio recorded in studio from about 100 speakers with little speaking style and text variation.
Such models struggle to synthesize speech with rich variation in emotion, voice, background noise, acoustic condition, and have not been tested on the abilities to generalize to tasks not explicitly trained on.

There had been a few attempts of using in-the-wild data such as CommonVoice~\citep{Ardila2019CommonVA}, Librispeech~\citep{Panayotov2015LibrispeechAA}, and LibriTTS~\citep{zen2019libritts} for training text-to-speech (TTS) models. It led to huge quality degradation compared to training on curated datasets~\citep{Hsu2018HierarchicalGM,Wang2021fairseqSA}. In particular, while in-the-wild data are generally of lower quality, the gap between synthesized and training speech is big compared to that of the models trained on curated speech~\citep{Wang2021fairseqSA}, which suggests that previous models terribly \textit{underfit} in-the-wild data.

This paper presents \vb{}, the most versatile text-conditioned speech generative model at scale. \vb{} is trained on a text-guided speech infilling task, where the goal is to generate masked speech given its surrounding audio and text transcript. This can be considered as a guided in-context learning problem, where audio style is inferred from the audio context and textual content is specified through transcript. \vb{} does not require any audio style labels (e.g., speaker, emotion, and noise), which differentiates \vb{} from the majority of prior work where such labels are used extensively. Prior work uses labels to make the mapping between input (text and audio style) and output (speech) more deterministic to reduce underfitting~\citep{Wang2021fairseqSA, Popov2021GradTTSAD}. We show that \vb{}'s text-guided speech infilling approach is much more scalable in terms of data while subsuming many common speech generative tasks.

In terms of modeling, \vb{} is a non-autoregressive (NAR) continuous normalizing flow (CNF) model~\citep{cnf}. Similar to diffusion models~\citep{Ho2020DenoisingDP}, 
CNFs model the transformation from a simple distribution to a complex data distribution ($p(\text{missing data} \mid \text{context})$), parameterized by a neural network. 
We train \vb{} with flow-matching~\citep{flow-matching}, a recently proposed method that enables efficient and scalable training of CNFs via a simple vector field regression loss. 
In contrast to auto-regressive models, \vb{} can consume context not only in the past but also in the future. Moreover, the number of flow steps can be controlled at inference time to flexibly trade off quality and runtime efficiency.

\vb{} is trained on 60K hours of English audiobooks and 50K hours of multilingual audiobooks in 6 languages for the mono and multilingual setups.
\vb{} achieves SOTA performance on mono-lingual/cross-lingual zero-shot TTS, speech denoising, speech editing, diverse speech sampling and an application to data creation for speech recognition. 
To tackle the lack of comparability due to the use of subjective metrics, this paper presents a series of metrics using public models to facilitate reproducible comparison and model development for speech generation studies. 

The contribution of this work can be summarized as follows:
\begin{enumerate}
    \item \vb{} represents a breakthrough in generative modeling for speech. By learning to solve a text-guided speech infilling task with large scale data, \vb{} can solve tasks it was not explicitly trained to accomplish via in-context learning. 
    \item \vb{} outperforms VALL-E and achieves a new SOTA English zero-shot TTS result (5.9\% $\rightarrow$ 1.9\% on word error rate (WER) and 0.580 $\rightarrow$ 0.681 on audio similarity).
    \item \vb{} is the first model that can perform high-quality cross-lingual zero-shot TTS across six languages. It does not use any style labels, pre-trained embedders, or multilingual samples. Compared to the prior cross-lingual SOTA YourTTS, \vb{} reduces the average WER from 10.9\% to 5.2\%, and improves audio similarity from 0.335 to 0.481.
    \item \vb{} is capable of infilling speech of any length and outperforms the prior SOTA A3T, on text guided denoising with -8.8\% WER, +0.450 similarity, and +0.80 mean opinion score.
    \item \vb{} can generate diverse and realistic speech. An ASR system can be trained solely on synthetic speech generated by \vb{}, resulting in only 0.4\%/1.7\% absolute WER increase on Librispeech test-other/test-clean compared to training on real data. In contrast, previous TTS models suffer from at least 18.2\%/44.5\% absolute WER increase.
\end{enumerate}
\section{Related Work}

\paragraph{Generative speech models}
Most speech generative models are task-specific and trained on different datasets. One common type of task is \textit{audio style conversion}, which aims to convert only a specific attribute while keeping other attributes the same. Voice conversion~\citep{Kameoka2018StarGANVCNM,LorenzoTrueba2018TheVC}, emotion conversion~\citep{robinson2019sequence,kreuk2021textless}, speech enhancement~\citep{xu2014regression,defossez2020real,Serr2022UniversalSE} belong to this category. 
Many of these models are supervised and trained on pairs of data that only differ in one attribute, for example, emotion~\citep{kreuk2021textless}. It is hard to obtain such data. Moreover, some attributes, such as speaking style, are hard to annotate. Hence, these models are often trained on small datasets and do not generalize well.

Controllable text-to-speech synthesis (TTS) is another common task, which aims to synthesize speech in a target audio style (e.g., voice, speaking style, recording environment) given text. While some styles like voice can be specified through labels~\citep{Kim2021ConditionalVA} or pre-trained embeddings like YourTTS~\citep{Casanova2021YourTTSTZ} and \citet{Jia2018TransferLF}; others like prosody are hard to annotate or embed. Previous studies~\citep{Wang2018StyleTU, Akuzawa2018ExpressiveSS, Hsu2018HierarchicalGM} tried to control them by learning a residual embedding. However, these models encode style in a low-dimensional space and impose an overly simple distribution of speech given text and residual embedding~\citep{Ren2020FastSpeech2F,Shen2017NaturalTS}. They cannot generate realistic noisy speech given a low dimensional vector, and performance degrades when conditioned on noisy references~\citep{Hsu2018HierarchicalGM}.

Infilling can be considered as another type of task. It aims to predict speech given context~\citep{Lakhotia2021OnGS, Borsos2022AudioLMAL} and optionally text guidance~\citep{Bai2022A3TAA, Borsos2022SpeechPainterTS, Wang2023NeuralCL}. Instead of learning an explicit embedding to control style, infilling models predict speech coherent to the context. In other words, these models perform in-context learning similar to Large Language Models (LLMs), which specifies the task (i.e., the desired style to convert) through context.
While this is a step toward building large scale generalist models using little explicit supervision, most prior work using text guidance still assumes a deterministic mapping from text and context to target~\citep{Bai2022A3TAA,Borsos2022SpeechPainterTS}, which is only realistic for very short segments. Hence, models with those assumptions could typically only infill segments up to 1 second~\citep{Borsos2022SpeechPainterTS}.
\vb{} is a text-guided infilling model, but it leverages the CNF model that can parameterize any distribution. Hence, \vb{} can infill speech of any length and can be trained on in-the-wild datasets with rich variation, and provide a general solution that subsumes many tasks in a text-guided fashion.

\paragraph{Large scale in-context learning models}
With the advancement in neural codec for speech~\citep{Hsu2021HuBERTSS, Defossez2022HighFN, Zeghidour2022SoundStreamAE}, many recent studies explore token-based language modeling for speech generation. The GSLM-family \citep{Lakhotia2021OnGS, Kharitonov2021TextFreePG, Nguyen2022GenerativeSD} are textless language models built upon HuBERT units~\citep{Hsu2021HuBERTSS} for speech continuation without using text. 
HuBERT units encode mostly content~\citep{polyak2021speech}, and the generated speech does not preserve the voice of the prompt. To tackle this, AudioLM~\citep{Borsos2022AudioLMAL} considers a cascaded approach which first generates HuBERT-like tokens and then predicts SoundStream~\citep{Zeghidour2022SoundStreamAE} tokens, a reconstruction based codec that preserves style. These models are not conditioned on text and are evaluated on spoken language modeling tasks.

VALL-E~\citep{Wang2023NeuralCL} is most related to \vb{}. It is a text conditioned LM trained on Encodec~\citep{Defossez2022HighFN} tokens (similar to SoundStream). Encodec tokenizes speech with a residual quantization layer, which encodes each frame with 8 codebooks at a 75Hz frame rate. The codebooks are ordered such that the first code contains the most information and so on. VALL-E has two modules. The first is an auto-regressive (AR) model that predicts the first code of each frame given text and the audio prompt. The second is an NAR model that predicts the remaining seven codebooks sequentially (all frames are predicted simultaneously when predicting each codebook).

VALL-E demonstrates state-of-the-art (SOTA) zero-shot TTS performance through in-context learning, where speech of the desired style is used as the prompt. The model considers the prompt as part of the whole utterance such that it generates the rest of the utterance containing the target text in the same audio style. \vb{} has several design advantages compared to VALL-E. 
1) \vb{} can use context both in the past and future, which is useful for editing where only a segment in the middle needs to be generated. 
2) \vb{} can generate speech much faster than VALL-E because flow-matching can produce high quality samples with less than ten NAR steps, while VALL-E requires one AR and seven NAR steps. 
3) \vb{} decouples duration and audio modeling, enabling finer grained alignment control.
4) \vb{} is compatible with any continuous features including Encodec embeddings.

NaturalSpeech2~\citep{shen2023naturalspeech} is another concurrent work that explores diffusion-style models for in-context speech generation. It adopts the latent diffusion framework~\citep{rombach2022high}, which encodes speech into latent features using an auto-encoder with a residual vector quantizer, and uses a diffusion model to generate latent features conditioned on text, (predicted) pitch, and a speech prompt. The predicted latent features are converted to waveform using the decoder of the same auto-encoder. 

NaturalSpeech2 differs from \vb{} in a few key aspects. 
1) it has an extra pitch predictor and conditions feature generation on pitch. 
2) NaturalSpeech2 predicts learned latent features while \vb{} predicts Mel spectrogram.
3) it adopts an asymmetric encoder for speech prompt and speech target, and uses two-stage cross-attention to query the prompt. Inference for NaturalSpeech2 is more efficient in terms of attention computation because the prompt is only encoded once and reused for every diffusion step, but it is unclear how this design affects infilling performance, which was not evaluated. 
4) the duration and the pitch predictor is a regression model which predicts only one duration and pitch for the same prompt and text.
5) NaturalSpeech2 always conditions on a speech prompt during inference, which does not allow diverse speech sampling demonstrated in \cref{sec:exp_speech_samp}.
6) \vb{} leverages flow-matching with optimal transport path which was shown to train and infer faster than diffusion paths, which is the objective NaturalSpeech2 adopts. \vb{} generates high quality samples with only 16 ODE steps while NaturalSpeech2 sets the diffusion steps to 150. The overall inference time for \vb{} is expected to be faster.
\section{Method}\label{sec:method}

\subsection{Background: Flow Matching with an optimal transport path}
Let $\mathbb{R}^d$ be the data space with data points $x \in \mathbb{R}^d$ drawn from some unknown distribution $q(x)$. 
Continuous Normalizing Flows (CNFs) \cite{cnf} are a family of generative models that learn the transformation from a simple prior distribution $p_0$ (e.g., normal distribution) to the data distribution $p_1 \approx q$. CNFs parameterize a time-dependent vector field $v_t: [0,1] \times \mathbb{R}^d \rightarrow \mathbb{R}^d$ that is used to construct a \textit{flow}: $\phi_t: [0,1] \times \mathbb{R}^d \rightarrow \mathbb{R}^d$ that pushes points from the prior towards the target distribution.  The relationship between a vector field and a flow is defined via the ordinary differential equation (ODE) as: 
\begin{equation}
    \dfrac{d}{dt} \phi_t(x) = v_t(\phi_t(x)); \quad \phi_0(x) = x.
\end{equation}

For a flow $\phi_t$, the \textit{probability path} (time-dependent probability density function) $p: [0,1] \times \mathbb{R}^d \rightarrow \mathbb{R}_{>0}$ can be derived via the change of variables formula: 
\begin{equation}
    \pt = p_0(\phi_t^{-1}(x)) \det \left[ \dfrac{\partial \phi_t^{-1}}{\partial x}(x) \right].
\end{equation}
To sample from $p_t(x)$, we first draw $x_0$ from $p_0$ and then solve the initial value problem (IVP) for $\phi_t(x_0)$ given $d\phi_t(x)/dt = v_t(\phi_t(x))$ and $\phi_0(x) = x_0$. We use $x_t$ and $\phi_t(x_0)$ interchangeably.

Let $p_t$ be a probability path and $u_t$ be the corresponding vector field that generates $p_t$. The vector field $v_t(x; \theta)$ parameterized by a neural network $\theta$ can be trained with the Flow Matching objective: 
\begin{equation}
    \mathcal{L}_{FM}(\theta) = \mathbb{E}_{t, \pt} || u_t(x) - v_t(x; \theta) ||^2,    
\end{equation}
where $t \sim \mathcal{U}[0,1]$ and $x \sim \pt$. While the objective appears simple, in practice we do not have the prior knowledge of $p_t$ or $v_t$, and cannot directly compute the loss or its gradient estimator.

Let $x_1$ be a random variable distributed according to data distribution $q$. \citet{flow-matching} first notes that a probability path $\pt$ can be constructed via a mixture of simpler \textit{conditional paths} $\cpt$ whose vector field $\cvf$ can be easily computed. To construct $\pt$, a conditional path is defined such that 1) $p_0(x \mid x_1) = p_0(x)$ and 2) $p_1(x \mid x_1) = \mathcal{N}(x \mid x_1, \sigma^2 I)$, a Gaussian distribution centered at $x_1$ with a sufficiently small $\sigma$ (typically $10^{-5}$). The marginal path is computed as $\int \cpt q(x_1)dx_1$, which closely approximates $q(x_1)$ at $t=1$. With that, \citep{flow-matching} presents the Conditional Flow Matching (CFM) objective, 
\begin{equation}\label{eq:cfm}
    \mathcal{L}_{CFM}(\theta) = \mathbb{E}_{t, q(x_1), \cpt}|| \cvf - v_t(x; \theta) ||^2.    
\end{equation}
It is proven that FM and CFM have identical gradients w.r.t. $\theta$. More importantly, one can easily draw samples from $\cpt$ and compute $\cvf$ to derive an unbiased gradient estimator.

The next question is \textit{how to choose a conditional flow.} A flow defines \textit{trajectories}, describing how each point moves between $p_0$ and $p_1$. Intuitively, a simpler trajectory (e.g., a straight line) can be learned faster and the IVP can be solved more accurately and efficiently.
\citet{flow-matching} presents a conditional flow called \textit{optimal transport (OT) path}, which has the form of $p_t(x \mid x_1) = \mathcal{N}(x \mid tx_1, (1-(1-\sigma_{min})t)^2 I)$ and $\cvf = \left( x_1 - (1 - \sigma_{\text{min}})x \right) / \left(1 - (1 - \sigma_{\text{min}})t \right)$. The flow is arguably simple because points move with a constant speed and direction. We adopt it for \vb{}

\citet{flow-matching} also presents another flow that recovers the path of diffusion models~\citep{song2019generative}, which is more complex than the OT path. We will present ablation studies comparing different paths (OT vs diffusion) and different objectives (CFM vs score-matching). Results show the superiority in performance and efficiency of CFM with OT path

\subsection{Problem formulation}

Given a dataset of transcribed speech $(x, y)$ where $x$ and $y$ denote an audio sample and its transcript, respectively, the goal is to build a single model that can perform many text-guided speech generation tasks through in-context learning.
We propose to train such a generative model on the \textit{text-guided speech infilling task}, which predicts a segment of speech given its surrounding audio and the complete text transcript. Let $m$ be a binary temporal mask which is of the same length as $x$, \footnote{``temporal'' refers to the physical time of the audio sample, not the time $t\in[0,1]$ of the flow in CNF.} 
and $\xmis = m \odot x$ and $\xctx = (1 - m) \odot x$ be the complementary masked versions of $x$. The generative model learns $p(\xmis \mid y, \xctx)$. In other words, $y$ and $\xctx$ are the context and $\xmis$ is the missing data.

\subsection{Model and Training}
\label{subsec:model_and_training}
Motivated by the need that some applications require fine-grained alignment control between speech and text, we decouple \vb{} into two components: an audio model and a duration model. 
Let $x = (x^1, x^2, \cdots, x^N)$ be an audio sample of $N$ frames, 
$y = (y^1, y^2, \cdots, y^M)$ be a text sequence of $M$ phones, 
and $l = (l^1, l^2, \cdots, l^M)$ be the per-phone duration where $l^j$ denotes how many audio frames $y^j$ correspond to and $\sum_{j=1}^M l^j = N$.
We further define $z = \text{\texttt{rep}}(y, l) = (z^1, z^2, \cdots, z^N)$ to be the frame-level phone transcript, which repeats each $y^j$ by $l^j$ times such that $z^i$ denotes the phone label of the audio frame $x^i$. 
For a pair of $(x, y)$, $l$ and $z$ can be estimated through \textit{forced alignment} using a speech recognition model.
The estimation of $q(\xmis \mid y, \xctx)$ is then broken down into the audio model $q(\xmis \mid z, \xctx)$ and the duration model $q(\lmis \mid y, \lctx)$, where $\lmis$ and $\lctx$ denote $l$ masked by $m'$ and $1 - m'$, and $m'$ is downsampled from $m$ based on $l$ where $m = \text{\texttt{rep}}(m', l)$

\begin{figure}[ht]
    \centering
    \includegraphics[width=0.75\linewidth]{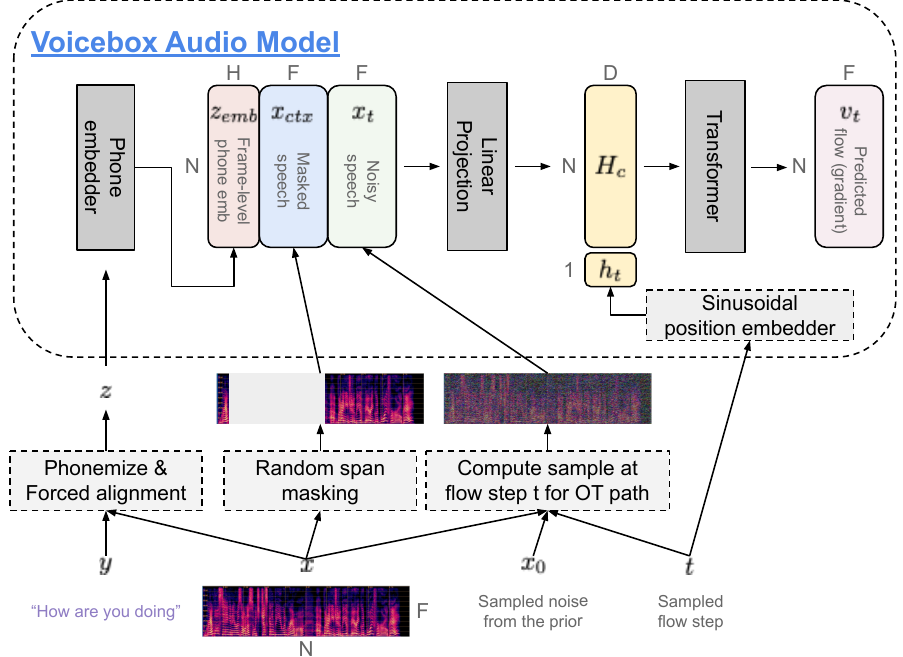}
    \caption{Illustration of the Voicebox audio model. The lower half illustrates how inputs are created during training. $N$ denotes the number of frames, $H$ for the phone embedding dimension, $F$ for the spectral feature dimension, and $D$ for the Transformer input dimension. Solid dark gray blocks denote trainable components, and light gray blocks with dashed border are frozen components or operations without trainable parameters.}
    \label{fig:audio_model}
\end{figure}

\paragraph{Audio Model (\cref{fig:audio_model})}
Given a context $z$ and $\xctx$ of length $N$, the distribution of $\xmis$ is highly stochastic especially when $\xmis$ has a large temporal span. Hence, we parameterize it with a CNF and train it using the flow matching objective with the optimal transport path. Audio $x$ is represented as an $80$-dimensional log Mel spectrogram ($x^i \in \mathbb{R}^{80}$) extracted at a 100Hz frame rate.\footnote{A Mel-spectrogram can be converted back to a waveform using a vocoder.} The audio context $\xctx^i = \mathbf{0}$ where $m^i = 1$ and $\xctx^i = x^i$ where $m^i = 0$. 

For simpler conditioning, we model the conditional distribution $q(x \mid z, \xctx)$ of all frames $x$ instead of only masked frames $\xmis$. A neural network is used to parameterize the conditional vector field $v_t(x_t, \xctx, z; \theta)$ that additionally takes $\xctx$ and $z$ as input. Note that $x_t$ is a sample at flow step $t$.

Given as input $\xctx \in \R{N \times F}$, $x_t \in \R{N \times F}$, phone sequence $z \in [K]^{N}$ with $K$ denoting the number of phone classes, and a time step $t \in [0,1]$, we employ a Transformer model to parameterize the vector field $v_t$. A lookup table, denoted as $L \in \R{K \times H}$, is used to embed the phone sequence $z$, resulting in the embedded sequence $z_{emb} \in \R{N \times H}$ where $z_{emb}^i = L(z^i)$ for $i \in {1,\dots,N}$.
Subsequently, the three sequences ($x_t$, $\xctx$, and $z_{emb}$) are concatenated frame-by-frame and projected by employing matrix $W_p \in \R{(2F+H) \times D}$, thereby obtaining the sequence $H_c \in \R{N \times D}$ where $D$ represents the embedding dimension of the Transformer model.

To embed the flow step, a sinusoidal positional encoding is applied to map $t \in [0,1]$ to $h_t \in \R{D}$. The sequence $\tilde{H}_c \in \R{(N+1) \times D}$, which serves as the input to the Transformer model, is derived by concatenating $H_c$ with the vector $h_t$ along the time dimension. 
Given the Transformer output $v_t(x_t, \xmis, z; \theta) \in \R{N \times F}$, which is the sub-sequence corresponding to $H_c$, the loss is computed as:
\begin{equation}\label{eq:unmasked-loss}
    \mathcal{L}_{\text{audio-CFM}}(\theta) = \mathbb{E}_{t, m, q(x, z), p_0(x_0)}|| u_t(x_t \mid x) - v_t(x_t, \xctx, z; \theta) ||^2,
\end{equation}
by reparameterizing \cref{eq:cfm}. During training, given an audio sample $x$ and a prior sample $x_0$, we have $x_t = (1 - (1 - \sigma_{\text{min}})t)x_0 + tx$ and $u_t(x_t \mid x) = x - (1 - \sigma_{min})x_0$. This function computes the loss on all frames, including those that are not masked and would not be required during inference. To divert the model's focus to masked frames, we present a masked version of $\mathcal{L}_{\text{audio-CFM}}$:
\begin{equation} \label{eq:masked-loss}
    \mathcal{L}_{\text{audio-CFM-m}}(\theta) = \mathbb{E}_{t, m, q(x, z), p_0(x_0)}|| m \odot \left( u_t(x_t \mid x) - v_t(x_t, \xctx, z; \theta) \right) ||^2,
\end{equation}
where the loss is only computed on masked frames. \cref{sec:app_train_obj} shows it leads to better results.

\paragraph{Duration model} 
We consider two solutions. The first one closely follows the audio model. It models $q(l \mid y, \lctx)$ via a conditional vector field which swaps $(x, \xctx, z)$ with $(l, \lctx, y)$ and accordingly for the flow, where $l, \lctx \in \R{M \times 1}$ and $y \in [K]^M$. The masked version of the CFM loss is used for training.
On the other hand, previous studies have shown that regression duration models can produce reasonable speech~\citep{Ren2020FastSpeech2F,fastpitch}. Hence we consider a second solution that regresses the masked duration $\lmis$ given the context duration $\lctx$ and phonetic transcript $y$. The same Transformer model is used, except that there are only two input sequences instead of three, and the time embedding is not used. The model is trained with an $L_1$ regression loss on masked phones:
\begin{equation}
    \mathcal{L}_{\text{dur-regr-m}}(\theta) = 
    \mathbb{E}_{m, q(l, y)} || m' \odot \left( \lmis - g(\lctx, y; \theta) \right) ||_1,
\end{equation}
where $g$ denotes the regression-based duration model. This is similar to the duration model used in FastSpeech2~\citep{Ren2020FastSpeech2F}, but with additional duration context $\lctx$ as input.

\subsection{Inference}\label{sec:method_inf}
\begin{figure}[h]
    \centering
    \includegraphics[width=0.7\linewidth]{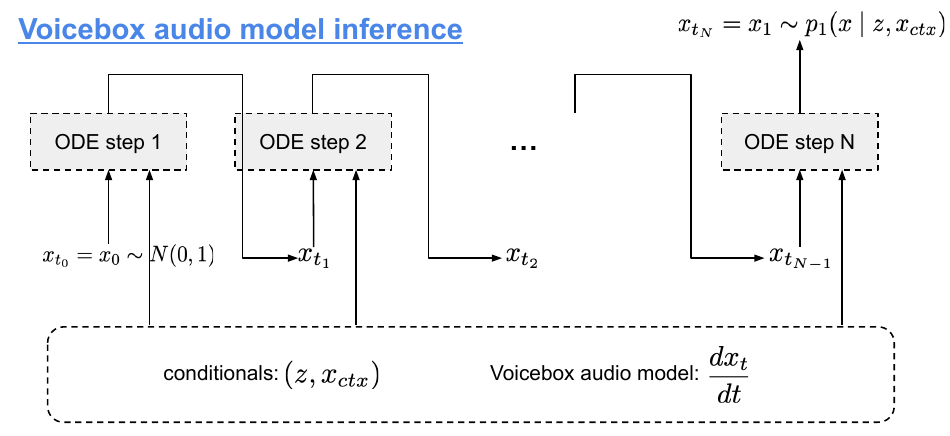}
    \caption{Inference is equivalent to solving an ODE with an initial condition $x_0$ sampled from the prior, a derivative $\frac{dx_t}{d_t}$ specified by the audio model, and conditional inputs $(z, x_{ctx})$. At each step, the ODE solver estimate $x_{t+1}$ given $t$, $x_t$ from the previous step, the audio model, and the conditional inputs. In the end, it produces $x_1$, which is a sample drawn from the learned distribution $p_1$.}
    \label{fig:audio_inference}
\end{figure}

To sample from the the learned audio distribution $p_1(x \mid z, \xctx)$, a noise $x_0$ is first sampled from $p_0$, and then an ODE solver is used to evaluate $\phi_1(x_0)$ given $d\phi_t(x_0)/dt = v_t(\phi_t(x_0), \xctx, z; \theta)$ and the initial condition $\phi_0(x_0) = x_0$. \cref{fig:audio_inference} provides an illustration of the process.

Intuitively, the ODE solver computes $\phi_1(x_0)$ by evaluating $v_t$ at multiple $t$ to approximate the integration from $t=0$ to $t=1$ given the initial condition $\phi_0(x_0) = x_0$. 
The number of function evaluation (NFE) is defined as how many times $d\phi_t(x_0)/dt$ is evaluated.
A higher NFE often leads to a more accurate solution of $\phi_1(x_0)$ at the cost of longer run time. 
This provides great flexibility for users to decide the trade-off between speed and accuracy. Moreover, we find that empirically \vb{} can generate very high quality speech with less than 10 NFEs, making it significantly faster than auto-regressive models.

\subsection{Classifier-Free Guidance}
Classifier guidance (CG)~\citep{dhariwal2021diffusion} is a technique used to trade off mode coverage and sample fidelity for diffusion models post training, similar to the effect of truncated or low-temperature sampling for generative adversarial networks~\citep{brock2018large} and discrete flow models~\citep{kingma2018glow}. It modifies the score estimate of a diffusion model to include the gradient of the log likelihood of an auxiliary classifier.
\citet{ho2022classifier} notes that CG approximates sampling from $p(x \mid c)p(c \mid x)^{\alpha}$ where $c$ is the conditioner, and this can be simulated without a classifier by mixing the score estimate of a conditional model and an unconditional model. The unconditional model can be jointly trained by dropping the conditioner $c$ with some probability, and the same model provides score estimates for both $p(x)$ and $p(x \mid c)$.

We extend the idea of classifier free guidance (CFG) to flow-matching models. The conditioner $c$ is equivalent to $(z, \xctx)$ for audio models and $(y, \lctx)$ for duration models, which is dropped with $p_{\text{uncond}}$ during training. During inference, the modified vector field $\tilde{v}_t$ for the audio model becomes
\begin{equation}
    \tilde{v}_t(w, \xmis, z; \theta) = (1+\alpha) \cdot v_t(w, \xctx, z; \theta) - \alpha \cdot v_t(w; \theta),
\end{equation} 
where $\alpha$ is the strength of the guidance, and $v_t(w; \theta)$ is obtained by dropping $\xctx$ and $z$.  We use $\alpha$ and $\alpha_{dur}$ for the CFG strengths for audio and duration model, respectively, which are selected based on empirical results.\footnote{Note that the computation is doubled for the same NFE when using CFG, because each evaluation of $\tilde{v}_t$ requires two forward passes of the model $v_t$.}

\subsection{Applications}
\begin{figure}[h]
    \centering
    \includegraphics[width=\linewidth]{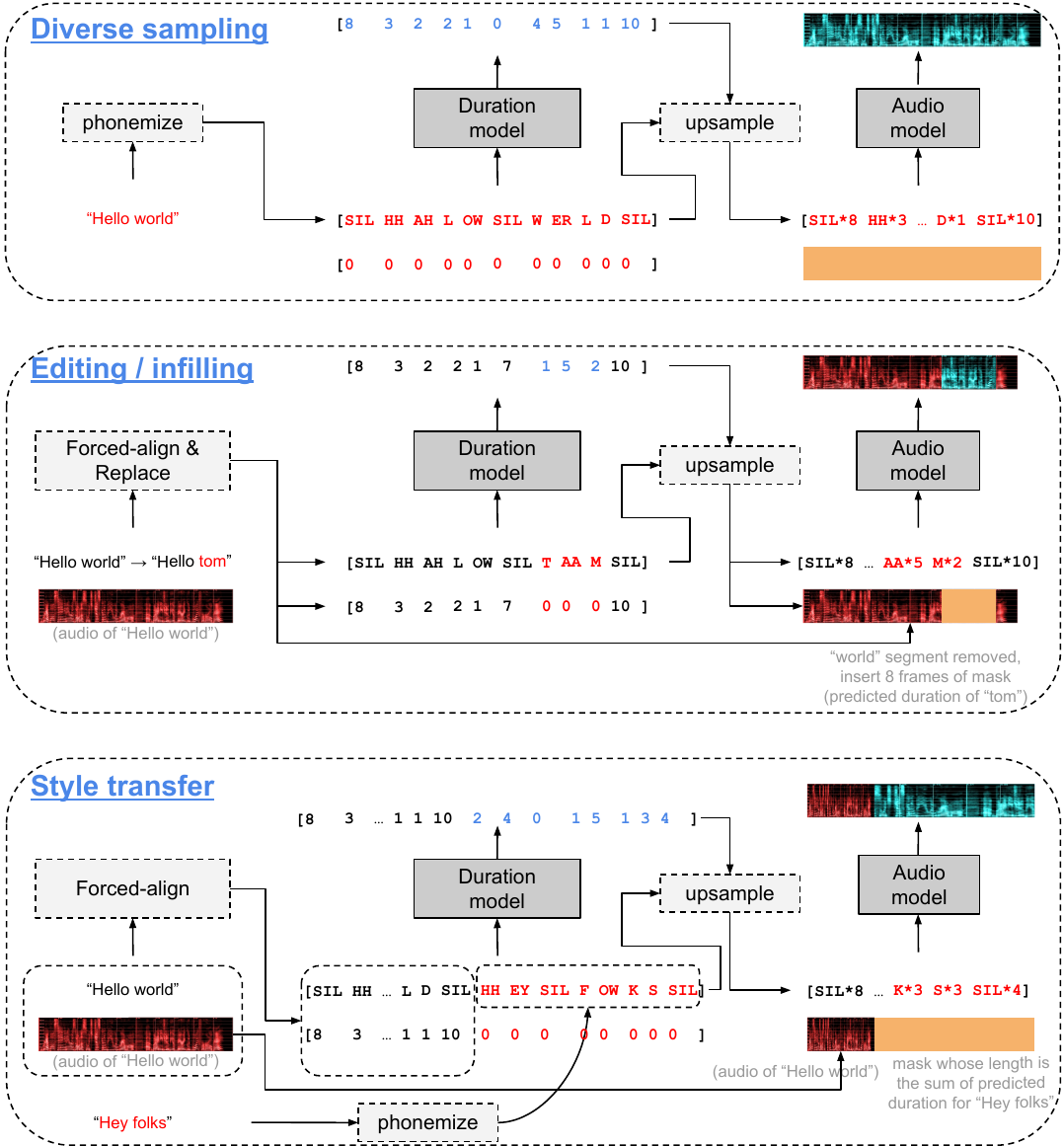}
    \caption{Detailed diagrams of diverse speech sampling, content editing, and style transfer. Text in red and blocks in orange at the input of a model denote segments to be predicted. Numbers in blue and spectrogram in cyan at the model output denote predicted duration and spectrogram.}
    \label{fig:app2}
\end{figure}

We demonstrate that \vb{} exhibits in-context learning abilities similar to LLMs by presenting a few examples of how to create context to perform tasks \vb{} was not explicitly trained on. These examples are also illustrated in \cref{fig:app,fig:app2}.

\paragraph{Zero-shot TTS \& alignment-preserved style transfer} Given a target text $\hat{y}$ and a transcribed reference audio $(x, y)$, zero-shot TTS aims to synthesize speech resembling the possibly unseen audio style of the reference. \vb{} performs the task by treating the reference audio and the target speech as one utterance where the target speech is masked. Let $l$ and $z$ be phone duration and frame-level transcript of $(x, y)$. The target duration $\hat{l}$ is sampled given the duration context $l$ and concatenated phone sequence $\text{\texttt{cat}}(y, \hat{y})$. The target speech $\hat{x}$ is then sampled given the context $x$ and concatenated frame-level phones $\text{\texttt{cat}}(z, \text{\texttt{rep}}(\hat{y}, \hat{l}))$.

\vb{} can also convert the audio style for speech $\bar{x}$ while preserving its alignment $\bar{z}$. This is useful for editing audio that is synchronized with other modalities such as video. Similar to zero-shot TTS, \vb{} can simply perform the task by sampling target speech $\hat{x}$ given the context $x$ and concatenated frame-level phones $\text{\texttt{cat}}(z, \bar{z})$.

\paragraph{Transient noise removal \& content editing} When recording speech, one might misspeak a few words or the recording my be interrupted by unexpected background noise. In these scenarios it is desired to just modify the problematic segment instead of re-recording the speech. 
\vb{} can perform transient noise removal through re-generating the noise corrupted segment given the original frame-level transcript and the surrounding clean audio. Specifically, given the frame-level transcript $z$ of a transcribed noisy speech $(x, y)$, a user creates a mask $m$ to indicate the noisy segment. The segment $\xhmis$ is then sampled given $z$ and $\xctx = (1-m) \odot x$.
The audio model would likely generate clean speech for $\xhmis$ because during training clean audio context co-occurs with clean target audio most of the time. The new audio $\hat{x} = \xhmis + \xctx$.

For content editing, let $\hat{y}$ be the new desired transcript with some words from the original transcript $y$ replaced, and $l$ be the original duration. A user first constructs $\lctx$ (of the same length as $\hat{y}$) by copying the lengths of phones that are not replaced from $l$, and set the lengths to $0$ for new phones. The duration of new phones $\lhmis$ is sampled given $\lctx$ and $\hat{y}$, and the new duration $\hat{l} = \lhmis + \lctx$. The new frame-level transcript is constructed with $\hat{z} = \text{\texttt{rep}}(\hat{y}, \hat{l})$. Similarly, the audio context $\xctx$ is of the same length as $\hat{z}$, and is created by filling frames mapped to unreplaced phones with the corresponding frames in $x$, and leaving those for new phones with $\mathbf{0}$. The frames for the new phones $\xhmis$ are sampled given $\hat{z}$ and $\xctx$. The edited speech is computed as $\hat{x} = \xhmis + \xctx$.

\paragraph{Diverse speech sampling \& alignment-preserved style shuffling} \vb{} can generate diverse speech samples by infilling the whole utterance. 
We first use the duration model to sample $\hat{l}$ given the phone transcript $\hat{y}$.  We then use the audio model to sample $\hat{x}$ 
given $\hat{z} = \text{\texttt{rep}}(\hat{y}, \hat{l})$.

Similar to style transfer, \vb{} can also shuffle the audio style while keeping the alignment by sampling $\hat{x}$ conditioning on the frame-level transcript $\bar{z}$ of the target speech clip $\bar{x}$.

\section{Metrics}

\vb{} formulates many speech generation tasks as text-guided in-context learning problems. The common goal of audio-conditioned tasks is to produce \textit{realistic} speech that is \textit{coherent} with the context and has the \textit{correct} textual content. For tasks not conditioned on audio context, it is desired to generate \textit{diverse and realistic} samples with distribution similar to training data with \textit{correct} content. 

Prior studies often adopt subjective metrics like mean opinion scores (MOS)~\citep{ribeiro2011crowdmos} which are not comparable across papers or even across studies in the same paper, because ratings can be biased by the quality of samples from other systems evaluated in the same trial. 
Some studies have also considered automatic quantitative metrics measuring signal-level similarity, such as mean cepstral distance (MCD)~\citep{kubichek1993mel,skerry2018towards} for speech synthesis and voice conversion, and signal-to-noise/distortion ratio (SNR/SDR)~\citep{le2019sdr} for speech enhancement. 
These metrics assume the output is deterministic given input, which is often ill-posed and unfairly penalizes generative models that produce realistic and valid samples. The caveats of signal level metrics have also been discussed in image generative modeling literature~\citep{saharia2022palette}.
In this paper, we advocate the following reproducible model-based perceptual metrics.

\paragraph{Correctness and intelligibility} This can be measured by the word error rate (\textbf{WER}) of the synthesized speech's transcription with respect to the input text, which has been adopted in prior work~\citep{Wang2018StyleTU}. Public automatic speech recognition (ASR) models are used for comparability. For English-only setups, we follow \citep{Wang2023NeuralCL} and use HuBERT-L~\citep{Hsu2021HuBERTSS} pre-trained on 60K hours of Librilight~\citep{Kahn2019LibriLightAB} and fine-tuned on 960 hours of Librispeech~\citep{Panayotov2015LibrispeechAA}. For multilingual setups we use the Whisper large-v2 model~\citep{Radford2022RobustSR}.

It should be noted that while a lower WER suggests that the generated speech is more intelligible by the model and contains more correct content, it does not necessarily imply the quality is better. 
Similarly, when generating diverse samples or when transferring to an audio style that is more expressive or more noisy, generated speech can be harder for an ASR model to recognize, which leads to a higher WER, which does not imply the sample is bad.

\paragraph{Coherence} This is measured by the similarity between the embedding of generated speech and that of the audio context, where different embedding models would reflect coherence of different attributes. VALL-E proposed to use WavLM-TDCNN speaker embedding model~\citep{Chen2021WavLMLS}, which maps an audio clip to a fixed dimensional vector, to measure voice similarity. We consider the same model to compare with VALL-E. 
In particular, VALL-E reports similarity with respect to \textit{resynthesized} audio context by its vocoder (Encodec-decoder), which we call \textbf{SIM-resyn (SIM-r)}. SIM-resyn is not comparable across models using different vocoders. Hence, we advocate for computing similarity against the original audio context, which we call \textbf{SIM-orig (SIM-o)}.

\paragraph{Diversity and quality} Fr\'{e}chet Inception Score (FID)~\citep{heusel2017gans} is widely adopted for image generation evaluations, which captures the similarity between generated and real images at the distribution level in some feature space. It fits a Gaussian distribution for real samples and one for generated samples in some feature space, and compute the Fr\'{e}chet distance between the two. A shorter distance implies the distributions are more similar and generally reflects \textit{both} higher sample quality and diversity.
We adapt the metric for speech by using self-supervised wav2vec 2.0 features~\citep{baevski2020wav2vec} and refer to it as Fr\'{e}chet Speech Distance (\textbf{FSD}). We verify its effectiveness in \cref{sec:app_fsd_ablation} along with alternative features.

As supplementary metrics, we include quality MOS (\textbf{QMOS}) for subjective audio quality evaluation, and similarity MOS (\textbf{SMOS}) for subjective audio similarity evaluation given pairs of prompt and system-generated audio clips. Both of which are in the scale of 1 to 5 with 5 being the best. 50 samples are evaluated for each system and 10 ratings are collected for each sample. Averaged ratings along with 95\% confidence interval are reported. For that, the CrowdMOS~\citet{ribeiro2011crowdmos} package was used with the recommended recipes for filtering outliers and inaccurate ratings. The MOS instructions can be found in \cref{sec:app_mos_instruct}.

To evaluate duration models, one can continue using the aforementioned metrics to gauge the end-to-end performance. Alternatively, we also present a few standalone metrics focusing on the duration model. Descriptions and results can be found in \cref{sec:app_dur_metrics,sec:app_dur_metrics_eval,sec:app_dur_e2e_eval}.

\section{Experiment}

\subsection{Setup}

\paragraph{Data} 
We train the English-only model on 60K hours ASR-transcribed English audiobooks and the multilingual model on 50K hours of multilingual audiobooks from six languages: English (En), French (Fr), German (De), Spanish (Es), Polish (Pl) and Portuguese (Pt). Following~\citep{Babu2022XLSR}, for a given upsampling factor $\beta$, we upsample low resource languages to mimic sampling batches from a multinomial distribution $p_s \sim\left(\frac{n_s}{N}\right)^\beta_{s=1, \ldots, S}$ where $S$ is the total number of languages, $n_s$ the number of pretraining hours of language $s$, and $N$ the total number of hours. We set $\beta=0.25$.

The two models are abbreviated as VB-En and VB-Multi. 
The Montreal Forced Aligner (MFA)~\citep{McAuliffe2017MontrealFA} is used to phonemize and force align the transcript based on the MFA phone set, which is a modified version of the international phonetic alphabet (IPA). Word position postfixes are added. Audio is represented as a 80-dimensional log Mel spectrogram and a HiFi-GAN vocoder trained on the same 60K hours of English speech is used to generate waveform. More details about phone representation, data transformation, and vocoder can be found in \cref{sec:app_phone_repr,sec:app_data_transform,sec:app_vocoder}.

\paragraph{Model}
Transformer~\citep{Vaswani2017AttentionIA} with convolutional positional embedding~\citep{baevski2020wav2vec} and symmetric bi-directional ALiBi self-attention bias~\citep{Press2021TrainST} are used for both the audio and the duration model. ALiBi bias for the flow step $x_t$ is set to 0. More details in Appendix \ref{sec:app_alibi}. The audio model has 24 layers, 16 attention heads, 1024/4096 embedding/feed-forward network (FFN) dimension, 330M parameters. We add skip connections connecting symmetric layers (first layer to last layer, second layer to second-to-last layer, etc.) in the style of the UNet architecture.  States are concatenated channel-wise and then combined using a linear layer.  The duration model has 8 heads, 512/2048 embedding/FFN dimensions, with 8/10 layers for English/multilingual setup (28M/34M parameters in total). All models are trained in FP16.

\paragraph{Training}
VB-En/VB-Multi audio models are trained for 500K/750K updates with an effective batch size of 240K frames. For training efficiency, audio length is capped at 1,600 frames and chunked randomly if the length exceeds this threshold. Duration models are trained for 600K updates with an effective batch size of 60K frames. The Adam~\citep{Kingma2014AdamAM} optimizer is used with a peak learning rate of 1e-4, linearly warmed up for 5K steps and linearly decays over the rest of training. For audio models, we clip the gradient norm to 0.2 for training stability. The audio/duration sequence is masked with $p_{\text{drop}}=0.3/0.2$, and otherwise a segment of $r\%$ sequence length is masked, where $r \sim \mathcal{U}[70,100]/\mathcal{U}[10,100]$. $p_{\text{uncond}}$ is set to $0.2$ for audio/duration models.

\paragraph{Inference}
The \texttt{torchdiffeq}~\citep{torchdiffeq} package is used, which implements both fixed and adaptive step ODE solvers. By default, the midpoint solver is used with a step size of 0.0625 (NFE=32). The regression duration model is used by default. Silence at both ends are trimmed to 0.1 second max.

\paragraph{Baselines} We consider three baselines: 
1) VALL-E~\citep{Wang2023NeuralCL}, SOTA for English zero-shot TTS trained on Librilight. 
2) YourTTS~\citep{Casanova2021YourTTSTZ}, SOTA multilingual (English, French, and Portuguese) zero-shot TTS model trained on VCTK, LibriTTS, TTS-Portugese~\citep{casanova2022tts}, and M-AILABS French. It is a flow-based model adapted from VITS~\citep{Kim2021ConditionalVA} using a pre-trained multilingual speaker embedder for voice conditioning.
3) A3T~\citep{Bai2022A3TAA}, SOTA for NAR speech editing and infilling trained with a regression loss on VCTK. 
We also consider Demucs~\citep{defossez2020real}, a SOTA speech enhancement model trained with regression and adversarial losses for denoising experiments. \cref{tab:baseline_capabilities} summarizes the tasks each baseline is capable of solving.

\begin{table}[ht]
    \caption{Comparing \vb{} with baselines on task capabilities. ``Sampling'' refers to the ability of generating diverse audio clips without conditioning on any audio. Through infilling, A3T and \vb{} can remove transient noise but not stationary background noise. VALL-E can only generate speech conditioning on the past context. Hence, the generated segment would only be coherent to the past context but will not have a smooth transition to the future context. With that, we label VALL-E as incapable of denoising or editing.}
    \label{tab:baseline_capabilities}
    \centering
    \begin{tabular}{l|ccccc}
        \toprule
        \textbf{Model} 
         & \textbf{ZS TTS} 
         & \textbf{Denoise} 
         & \textbf{Partial Edit} 
         & \textbf{Sampling} \\
        \midrule
        VALL-E & \gcell\cmark & \rcell\xmark & \rcell\xmark & \gcell\cmark \\
        YourTTS & \gcell\cmark & \rcell\xmark & \rcell\xmark & \gcell\cmark \\
        A3T & \gcell\cmark & \gcell\cmark (short) & \gcell\cmark & \rcell\xmark \\
        Demucs & \rcell\xmark & \gcell\cmark & \rcell\xmark & \rcell\xmark \\ 
        \midrule
        Voicebox & \gcell\cmark & \gcell\cmark (short) & \gcell\cmark & \gcell\cmark \\
        \bottomrule
    \end{tabular}
\end{table}

\subsection{Monolingual zero-shot TTS}\label{sec:exp_zs_tts}
\cref{tab:tts} presents the zero-shot TTS results of the English model VB-En. Following \citep{Wang2023NeuralCL}, the test set is constructed by selecting 4 to 10 second long samples from Librispeech test-clean. We consider \textit{cross-sentence} prompting where a 3 second clip from another sample of the same speaker is used as audio context, and \textit{continuation} where the first 3 seconds of each utterance is used.

We ran subjective MOS studies comparing ground truth, YourTTS, and \vb{}. A3T is not included because of the bad performance and VALL-E is not included because the model is not available. 
\vb{} outperforms all baselines on all metrics in both cases. In particular, \vb{} transfers style much more effectively (+0.101/+0.108 SIM-r on cross-sentence/continuation) than VALL-E, and the gap is even bigger when compared against raw audio (+0.141 SIM-o on continuation). MOS studies also confirm the quality and similarity of \vb{} are subjectively better than YourTTS.

\begin{table}[ht]
    \caption{English zero-shot TTS results on filtered LS test-clean. "-" results are not available. We obtain VALL-E continuation SIM result through communication with the authors.}
    \label{tab:tts}
    \centering
    \begin{tabular}{l|ccccc}
    \toprule
    \textbf{Model} & \textbf{WER} & \textbf{SIM-o} & \textbf{SIM-r} & \textbf{QMOS} & \textbf{SMOS}  \\
    \midrule
    Ground truth & 2.2 & 0.754 & n/a & 3.98{\tiny $\pm$ 0.14} & 4.01{\tiny$\pm$0.09} \\
    \midrule
    \multicolumn{5}{l}{\textit{cross-sentence}} \\
    A3T& 63.3	& 0.046 & 0.146 & - & - \\
    YourTTS & 7.7 & 0.337 & n/a  & 3.27{\tiny $\pm$ 0.13} & 3.19{\tiny$\pm$0.14} \\
    VALL-E & 5.9 & - & 0.580 & - & -  \\
    VB-En & 1.9 & 0.662 & 0.681 & 3.78{\tiny $\pm$ 0.10} & 3.71{\tiny$\pm$0.11} \\
    \midrule
    \multicolumn{5}{l}{\textit{continuation}} \\
    A3T & 18.7 & 0.058 & 0.144 & - & -  \\
    VALL-E & 3.8 & 0.452$^*$ & 0.508 & - & -  \\
    \colorours
    VB-En ($\alpha=0.7$) & 2.0 & 0.593 & 0.616 & - & -  \\ 
    \bottomrule
    \end{tabular}
\end{table}

\subsection{Cross-lingual zero-shot TTS}\label{sec:exp_cross_zs_tts}
\cref{tab:ttsx,tab:ttsx_smos} presents cross-lingual zero-shot TTS results, where the audio context and the target text are in different languages. Note that VB-Multi is \textit{not} trained on any sample with multiple languages in an utterance spoken by the same speaker. The test set is constructed using filtered MLS test split described in \cref{sec:app_mls_filter}. For each target text, we sample one 3-second long audio context from each language, which creates 36 language transfer directions in total.

\vb{} yields better performance than YourTTS everywhere. Specifically, on En/Fr/Pt which YourTTS supports, \vb{} obtains 3.1\%/5.9\%/8.1\% lower WERs and 0.136/0.141/0.160 higher similarity averaged across audio context in six languages. The average audio similarity MOS is 0.59 (3.89 vs 3.30) higher for \vb{} and the average quality MOS is 0.27 (3.50 vs. 3.23) higher.

\begin{table}[ht]
    \caption{Multilingual zero-shot TTS results on filtered MLS test sets. GT/YT/VB-Multi refers to ground truth/YourTTS/multilingual \vb{}. ``Ref'' column shows the audio context language.}
    \label{tab:ttsx}
    \centering
    \resizebox{\linewidth}{!}{
    \begin{tabular}{cc|cc|cc|cc|cc|cc|cc}
    \toprule
    \multirow{2}{*}{\textbf{}} & 
    \multirow{2}{*}{\textbf{Ref}} & 
    \multicolumn{2}{c|}{De} &  
    \multicolumn{2}{c|}{En} & 
    \multicolumn{2}{c|}{Es} &  
    \multicolumn{2}{c|}{Fr} &  
    \multicolumn{2}{c|}{Pl} &  
    \multicolumn{2}{c}{Pt} \\
    & 
    & \textbf{WER} & \textbf{SIM-o}
    & \textbf{WER} & \textbf{SIM-o}
    & \textbf{WER} & \textbf{SIM-o}
    & \textbf{WER} & \textbf{SIM-o}
    & \textbf{WER} & \textbf{SIM-o}
    & \textbf{WER} & \textbf{SIM-o}\\
    \midrule
    GT & - & 5.9 & 0.725 & 5.0 & 0.636 & 4.1 & 0.729 & 5.2 & 0.714 & 4.9 &  0.743 & 5.8 & 0.725 \\
    \midrule
    \multirow{7}{*}{YT}
    &De& n/a &n/a& 7.3 & 0.373 & n/a &n/a& 11.3 & 0.361 & n/a &n/a& 13.7 & 0.263 \\
    &En& n/a &n/a& 7.0 & 0.403 & n/a &n/a& 11.4 & 0.298 & n/a &n/a& 14.1 & 0.234 \\
    &Es& n/a &n/a& 7.6 & 0.327 & n/a &n/a& 11.6 & 0.316 & n/a &n/a& 13.5 & 0.256 \\
    &Fr& n/a &n/a& 7.6 & 0.363 & n/a &n/a& 10.7 & 0.459 & n/a &n/a& 13.1 & 0.299 \\
    &Pl& n/a &n/a& 7.8 & 0.349 & n/a &n/a& 11.8 & 0.370 & n/a &n/a& 15.1 & 0.308 \\
    &Pt& n/a &n/a& 7.6 & 0.322 & n/a &n/a& 11.8 & 0.297 & n/a &n/a& 13.6 & 0.436 \\
    \cmidrule{2-14}
    & \bf{AVG} & n/a &n/a & 7.5 & 0.356 & n/a &n/a & 11.4 & 0.350 & n/a &n/a & 13.9 & 0.299 \\

    \midrule
    \colorours & De & 4.8 & 0.632 & 4.8 & 0.522 & 3.6 & 0.442 & 5.3 & 0.489 & 5.5 & 0.449 & 5.4 & 0.420 \\
    \colorours & En & 5.9 & 0.435 & 4.2 & 0.535 & 4.1 & 0.423 & 6.8 & 0.423 & 8.3 & 0.402 & 7.6 & 0.385 \\
    \colorours & Es & 4.9 & 0.460 & 4.3 & 0.479 & 3.6 & 0.613 & 5.3 & 0.473 & 5.2 & 0.436 & 5.4 & 0.435 \\
    \colorours & Fr & 4.9 & 0.476 & 4.3 & 0.485 & 3.7 & 0.479 & 5.1 & 0.602 & 4.8 & 0.408 & 5.4 & 0.418 \\
    \colorours & Pl & 4.7 & 0.491 & 3.8 & 0.503 & 3.5 & 0.528 & 5.1 & 0.503 & 4.0 & 0.641 & 4.9 & 0.476 \\
    \colorours & Pt & 4.9 & 0.422 & 4.6 & 0.426 & 3.7 & 0.476 & 5.5 & 0.453 & 4.8 & 0.406 & 5.2 & 0.620 \\
    \colorours \cmidrule{2-14}
    \colorours 
    \multirow{-7}{*}{
        \begin{tabular}{@{}c@{}}VB-Multi \\ ($\alpha=1.0$)\end{tabular}
    } &  \bf{AVG} & 5.0 & 0.486 & 4.4 & 0.492 & 3.7 & 0.494 & 5.5 & 0.491 & 5.5 & 0.457 & 5.7 & 0.459 \\
    \bottomrule
    \end{tabular}
    }
\end{table}

\begin{table}[ht]
    \centering
    \caption{Multilingual zero-shot TTS SMOS/QMOS results on filtered MLS English test set with prompts in different languages. YT/VB-Multi refers to YourTTS/multilingual \vb{}. ``Ref'' shows the audio context language.}
    \label{tab:ttsx_smos}
    \begin{tabular}{c|cccccc}
    \toprule
    
    & Ref=De & Ref=En & Ref=Es & Ref=Fr & Ref=Pl & Ref=Pt \\
    \midrule
    & \multicolumn{6}{c}{\textbf{SMOS} (target text = En)} \\
    YT                      & 3.26{\tiny$\pm$0.11} & 3.24{\tiny$\pm$0.11} & 3.22{\tiny$\pm$0.12} 
                            & 3.48{\tiny$\pm$0.10} & 3.26{\tiny$\pm$0.09} & 3.38{\tiny$\pm$0.11} \\
    \colorours                            
    VB-Multi ($\alpha=1.0$) & 3.89{\tiny$\pm$0.10} & 3.93{\tiny$\pm$0.08} & 3.84{\tiny$\pm$0.10} 
                            & 3.92{\tiny$\pm$0.09} & 3.81{\tiny$\pm$0.08} & 3.96{\tiny$\pm$0.09} \\
    \midrule 
    & \multicolumn{6}{c}{\textbf{QMOS} (target text = En)} \\
    YT & 3.29{\tiny$\pm$0.12} & 3.17{\tiny$\pm$0.13} & 3.29{\tiny$\pm$0.12} & 3.08{\tiny$\pm$0.12} & 3.35{\tiny$\pm$0.12} & 3.21{\tiny$\pm$0.12} \\
    \colorours
    VB-Multi ($\alpha=1.0$) & 3.67{\tiny$\pm$0.09} & 3.48{\tiny$\pm$0.09} & 3.45{\tiny$\pm$0.11} & 3.31{\tiny$\pm$0.12} & 3.75{\tiny$\pm$0.11} & 3.35{\tiny$\pm$0.13} \\
    \bottomrule
    \end{tabular}
\end{table}

\subsection{Transient noise removal}\label{sec:exp_noise_removal}
We construct a noisy test set by mixing the filtered Librispeech test-clean from \cref{sec:exp_zs_tts} with non-speech noise such that it overlaps with 50\% of the duration at a -10dB signal-to-noise ratio. Note that for infilling models like A3T and \vb{}, the type and the SNR of transient noise would not affect the performance, because the corrupted segment is entirely masked and speech is re-generated independent of the corrupted segment.
Results of additional conditions can be found in \cref{sec:app_infill}.

\cref{tab:infill} presents the results comparing \vb{} with A3T and Demucs. It should be noted that A3T and \vb{} utilize transcript and location of the noise while Demucs does not. Nevertheless the goal of the study is to present a new paradigm and show \vb{} can perform denoising without being explicitly trained. 
Compared to the baselines, \vb{} generates samples that are much more intelligible (2.0\% WER), more similar to the clean parts of the audio (0.612 SIM-o), and of higher quality (3.87 MOS) in this challenging noise condition.
A3T is better than Demucs on intelligibilty and quality, but the infilled speech is not coherent because it is only trained on VCTK and cannot generalize to new audio styles. 

\begin{table}[h]
    \caption{Transient noise removal where noise overlaps with 50\% of the speech at a -10dB SNR.}
    
    \label{tab:infill}
    \centering
    \begin{tabular}{l|ccc}
    \toprule
    \textbf{Model}
    & \textbf{WER} & \textbf{SIM-o} & \textbf{QMOS} \\
    \midrule
    Clean speech & 2.2 & 0.687 & 4.07{\tiny$\pm$0.15} \\
    Noisy speech & 41.2 & 0.287 & 2.50{\tiny$\pm$0.15} \\
    \midrule
    Demucs & 32.5 & 0.368 & 2.86{\tiny$\pm$0.17} \\
    A3T & 11.5 & 0.148 & 3.10{\tiny$\pm$0.15} \\
    \colorours
    VB-En ($\alpha=0.7$) & 2.0 & 0.612 & 3.87{\tiny$\pm$0.17} \\
    \bottomrule
    \end{tabular}
\end{table}

\subsection{Diverse speech sampling and application to ASR data generation}\label{sec:exp_speech_samp}
\cref{tab:samp} compares the ability to generate diverse samples for Librispeech test-other text. We consider English \vb{} (VB-En) with regression (regr) or flow-matching (FM) duration models. VITS-VCTK additionally conditions on a speaker ID, which we randomly sample for each sentence. YourTTS conditions on text and a reference audio, which we draw from the LS train splits.

Qualitatively, A3T generates the same robotic voice when not conditioned on audio context and VITS-LJ generates high quality but a single voice, hence both yield high FSD (bad quality or diversity) but VITS-LJ has a low WER. VITS-VCTK improves the voice diversity and FSD and YourTTS further advances it as it is trained on more speakers.
\vb{} models (with different duration samplers) outperform the baseline on FSD by large margins, showing \vb{}'s ability to produce realistic and diverse samples whose distribution is close to the training data. Among them, the FM duration model creates more varying speaking styles compared to the regression one which ASR may struggle more to recognize. \vb{} even yields lower FSDs than the real samples from the Librispeech test-other split, because the latter contains only tens of speakers and the diversity is limited.

\begin{table}[ht]
\begin{minipage}{0.44\linewidth}
    \caption{Diverse speech generation from LS test-other text.}\label{tab:samp}
    \centering
    \resizebox{\linewidth}{!}{
        \begin{tabular}{l|cc}
        \toprule
        \textbf{Model} & \textbf{WER} & \textbf{FSD} \\
        \midrule
        Ground truth & 4.3 & 171.1 \\
        \midrule
        \multicolumn{3}{l}{\textit{require additional input}} \\
        VITS-VCTK & 10.6 & 306.6 \\
        YourTTS (ref=LS train) & 9.0 & 277.9 \\
        \midrule
        \multicolumn{3}{l}{\textit{text-only}} \\
        A3T & 37.9 & 373.0 \\
        VITS-LJ & 5.6 & 344.2 \\
        \colorours
        VB-En ($\alpha=0$, dur=regr) & 3.1 & 155.7 \\
        \colorours
        VB-En ($\alpha=0$, dur=FM, $\alpha_{dur}=0$) & 5.6 & 159.8 \\
        \bottomrule
        \end{tabular}
    }
\end{minipage}
\hfill
\begin{minipage}{0.51\linewidth}
    \caption{Performance of ASR models trained on real or synthetic speech, tested on \textit{real} speech and decoded with or without a 4-gram language model.}\label{tab:asr}
    \centering
    \resizebox{\linewidth}{!}{
    \begin{tabular}{l|cccc}
    \toprule
    & \multicolumn{4}{c}{\textbf{WER on real data}} \\
    & \multicolumn{2}{c}{\textbf{No LM}} 
    & \multicolumn{2}{c}{\textbf{4-gram LM}} \\
    \textbf{ASR training data} 
    &  test-c & test-o
    &  test-c & test-o \\
    \midrule
    Real audio (100hr) & 9.0 & 21.5 & 6.1 & 16.2 \\
    Real audio (960hr) & 2.6 & 6.3  & 2.2 & 5.0  \\
    \midrule
    VITS-LJ & 58.0 & 81.2 & 51.6 & 78.1 \\
    VITS-VCTK & 33.8 & 55.5 & 30.2 & 53.1 \\
    YourTTS (ref=LS train) & 25.0 & 54.6 & 20.4 & 51.2 \\
    \colorours
    VB-En ($\alpha=0$, dur=regr) & 7.1 & 17.6 & 6.5 & 14.6 \\
    \colorours
    VB-En ($\alpha=0$, dur=FM, $\alpha_{dur}=0$) & 3.1 & 8.3 & 2.6 & 6.7 \\
    \bottomrule
    \end{tabular}
    }
\end{minipage}
\end{table}

We next train an ASR model using \textit{only synthetic speech} and evaluate it on \textit{real speech}, which has not been successful before because synthetic data were not realistic and representative enough. \cref{tab:asr} compares real and synthetic data from \vb{} and three baseline models. Each TTS model generates one sample per text from the Librispeech training set, resulting in 281K utterances per system. For real data, we consider train-960 and train-clean-100. Details about the ASR model and training configurations are in \cref{sec:app_asr_details}. 

The results are highly correlated with the FSD scores of synthetic data. VITS-LJ gives the worst results, because the synthetic speech has only one voice. YourTTS performs better on test-clean, but is similar to VITS-VCTK on test-other. It suggests that while YourTTS is trained on more voices and can generate speech with higher voice diversity, it still fails to produce realistic noisy speech and hence the resulting ASR model still underperforms on test-other.

Both \vb{} variants beat the baseline by a large margin. In particular, \vb{} generates more diverse speech when using the FM duration model, which leads to a better ASR system when used for training. Compared to the baselines, the ASR model trained on \vb{} data with FM duration model reduces WERs by over 85\% and only lags behind real data by 0.4\% and 1.7\% absolute.

\subsection{Inference efficiency versus performance}\label{sec:eff}
\begin{figure}[ht]
\centering
\begin{subfigure}[t]{\columnwidth}
    \centering
    \includegraphics[width=0.78\columnwidth]{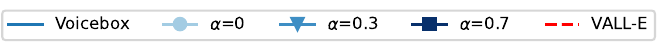}
\end{subfigure}

\begin{subfigure}[t]{.32\textwidth}
    \centering
    \includegraphics[width=\textwidth]{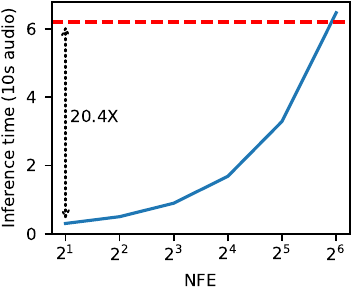}
    \caption{NFE vs time (w/o CFG)}
    \label{fig:nfe_vs_inference_time}
\end{subfigure}
\hfill
\begin{subfigure}[t]{0.32\columnwidth}
    \centering
    \includegraphics[width=\linewidth]{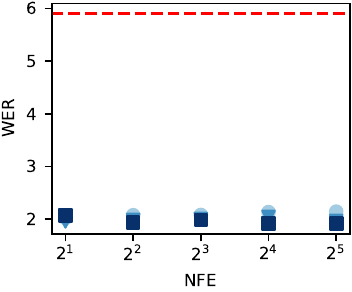}
    \caption{NFE vs WER (Zero-shot TTS)} \label{fig:nfe_vs_wer_zero_tts}
\end{subfigure}
\hfill
\begin{subfigure}[t]{.32\textwidth}
    \centering
    \includegraphics[width=\textwidth]{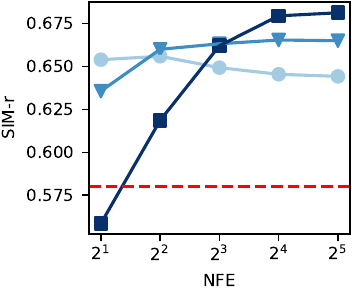}
    \caption{NFE vs SIM-r (Zero-shot TTS)}
    \label{fig:nfe_vs_similarity}
\end{subfigure}
\vspace{.25cm}

\begin{subfigure}[t]{0.49\columnwidth}
    \centering
    \includegraphics[width=0.64\linewidth]{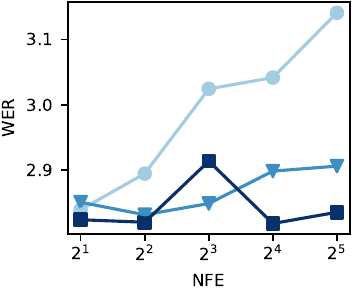}
    \caption{NFE vs WER (Diverse speech sampling)} \label{fig:nfe_vs_wer_vbox_tts}
\end{subfigure}
\hfill
\begin{subfigure}[t]{0.49\columnwidth}
    \centering
    \includegraphics[width=0.64\textwidth]{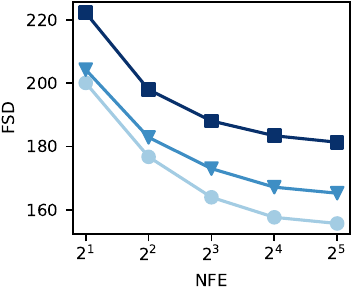}
    \caption{NFE vs FSD (Diverse speech sampling)}\label{fig:nfe_vs_fsd}
\end{subfigure}
\caption{Trade-off between NFE and different metrics. Inference time will be doubled with CFG.}
\label{fig:inference_efficiency_plots}
\end{figure}

We examine the trade-off between the metrics of interest (WER, SIM, FSD) for different settings of guidance strength ($\alpha$) and NFE specified by the user. \cref{fig:nfe_vs_inference_time} shows the \vb{} inference time to generate an audio sample of 10 seconds (including vocoding and predicting duration) as NFE varies and compares that to VALL-E.\footnote{Re-implemented and confirmed with the authors that our re-implementation is faster (6.2 vs 10 seconds).} 
For NFE=2 without CFG, \vb{} takes about $0.31$ seconds, about 20 times faster than VALL-E. At NFE=64, \vb{} is only $4\%$ slower than VALL-E.

Next, we study the cross-sentence setup of \cref{sec:exp_zs_tts} to analyze the impact on WER and SIM-r. We find that for all settings \vb{} has better WER than VALL-E. WER remains stable with mean of $2.0$ and variance of $0.005$ as shown in \cref{fig:nfe_vs_wer_zero_tts}.
\cref{fig:nfe_vs_similarity} shows that, in the case of SIM-r, lower classifier guidance strength values ($\alpha=0 \text{ or } 0.3$) produce higher speaker similarity when operating in a lower NFE regime ($\leq$ 4). However, starting from NFE=8, a higher classifier guidance strength improves speaker similarity.

Finally, in \cref{fig:nfe_vs_fsd} we examine FSD by generating samples for Librispeech test-other text. We find that lower classifier guidance strength produces lower FSD scores and more diverse samples. Increasing the NFE for each setting improves FSD.
\cref{fig:nfe_vs_wer_vbox_tts} shows the WER of the same test case. We find that for $\alpha=0$, WER increases slightly from $2.8$ to $3.1$ as NFE goes from $2$ to $32$. For a larger classifier guidance strength, WER remains more stable.
Through FSD and subjective listening, we discovered that a lower NFE leads to generating less diverse samples especially when the guidance weight is lower ($\alpha=0 \text{ or } 0.3$). Although those samples are of lower quality, they are easier for the ASR model to recognize because they tend not contain extreme audio styles like whispering or high background noise. As a result, WERs are lower.

\subsection{How context length affects monolingual and cross-lingual zero-shot TTS}\label{sec:app_ctx_leng}
\begin{figure*}[h]
    \centering
    \begin{subfigure}[t]{0.47\columnwidth}
        \includegraphics[width=\linewidth]{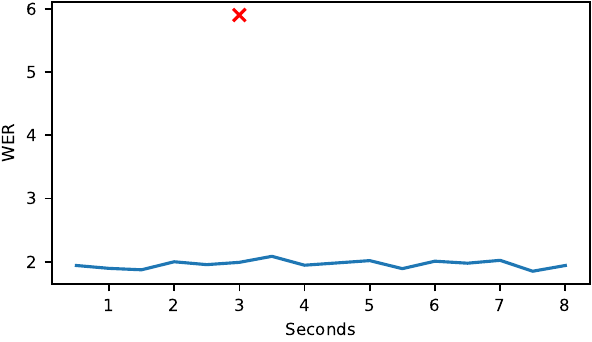}
        \caption{WER} \label{trnn:fig:prompt_wer}
    \end{subfigure}
    \begin{subfigure}[t]{0.04\columnwidth}
    \end{subfigure}
    \begin{subfigure}[t]{0.47\columnwidth}
        \includegraphics[width=\linewidth]{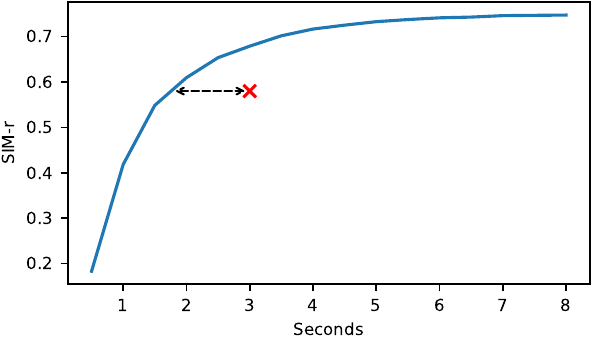}
        \caption{Speaker Similarity} \label{trnn:fig:prompt_spk}
    \end{subfigure}
    \caption{
        WER and SIM-r as a function of prompt audio time in seconds for the Zero-shot TTS task \ref{sec:exp_zs_tts}. Audio is generated using classifier-free guidance strength ($\alpha$) of $0.7$ and midpoint ODE solver with a NFE of $32$. The blue line is for \vb{} and the red star is VALLE at $3$ seconds. The speaker similarity (SIM-r) remains same for longer prompts (up to 10s).  
    }
    \label{fig:prompt_efficiency}
\end{figure*}

\boldparagraph{Monolingual} 
For in-context zero-shot TTS in Section~\ref{sec:exp_zs_tts}, we used $3.0$ seconds of prompt audio.  Here we examine how WER / SIM-r vary with different amounts of prompt audio using duration from regression duration model for the target text. If the desired prompt is longer than the available audio, the shorter audio is used as the prompt. Results are shown in Figure~\ref{fig:prompt_efficiency}.  As expected, WER mildly decreases and SIM-r grows quickly and flattens with longer audio prompts.  Comparing against VALL-E, \vb{} is more efficient at leveraging an audio prompt, achieving the same speaker similarity as VALL-E with roughly two thirds the input audio. 

\begin{figure}[h]
\centering
\begin{subfigure}[h]{0.8\columnwidth}
    \centering
    \includegraphics[width=0.8\columnwidth]{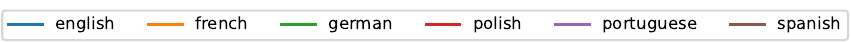}
\end{subfigure}
\hfill
\begin{subfigure}[h]{.32\textwidth}
	\centering
	\includegraphics[width=\textwidth]{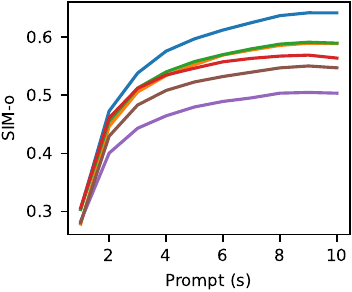}
	\caption{{English}}
	\label{fig:prompt_vs_similarity_english}
\end{subfigure}
\hfill
\begin{subfigure}[h]{.32\textwidth}
	\centering
	\includegraphics[width=\textwidth]{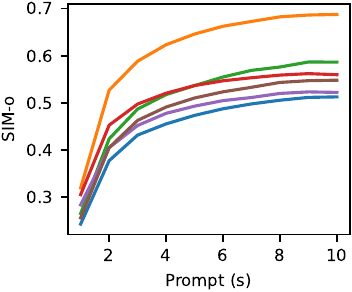}
	\caption{{French}}
	\label{fig:prompt_vs_similarity_french}
\end{subfigure}
\hfill
\begin{subfigure}[h]{.32\textwidth}
	\centering
	\includegraphics[width=\textwidth]{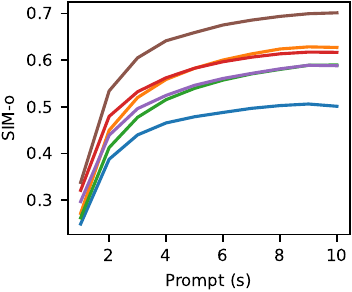}
	\caption{{Spanish}}
	\label{fig:prompt_vs_similarity_spanish}
\end{subfigure}
\hfill
\begin{subfigure}[h]{.32\textwidth}
	\centering
	\includegraphics[width=\textwidth]{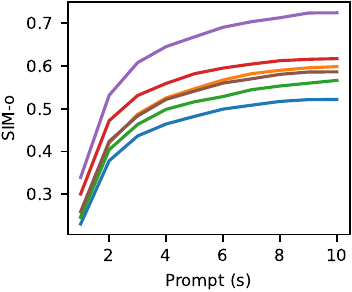}
	\caption{{Portuguese}}
	\label{fig:prompt_vs_similarity_portuguese}
\end{subfigure}
\hfill
\begin{subfigure}[h]{.32\textwidth}
	\centering
	\includegraphics[width=\textwidth]{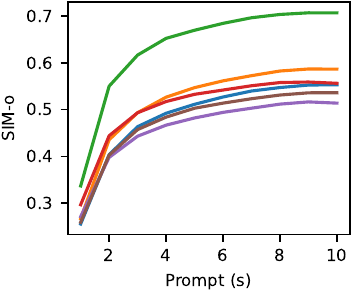}
	\caption{{German}}
	\label{fig:prompt_vs_similarity_german}
\end{subfigure}
\hfill
\begin{subfigure}[h]{.32\textwidth}
	\centering
	\includegraphics[width=\textwidth]{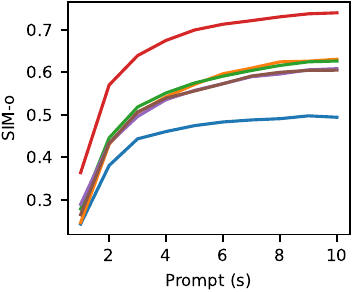}
	\caption{{Polish}}
	\label{fig:prompt_vs_similarity_polish}
\end{subfigure}
\caption{Each subplot considers one of the six target language and shows SIM-o (speaker similarity) as a function of prompt audio duration in seconds for cross-lingual style transfer from different source language. We set the classifier-free guidance strength ($\alpha$) to $1.0$ and use midpoint ODE solver with a NFE of $32$.}
\label{fig:prompt_vs_similarity_mls}
\end{figure}

\begin{figure}[h]
\centering
\begin{subfigure}[h]{0.8\columnwidth}
    \centering
    \includegraphics[width=0.8\columnwidth]{figs/mls_prompts/mls_legend.pdf}
\end{subfigure}
\hfill
\begin{subfigure}[h]{.32\textwidth}
	\centering
	\includegraphics[width=\textwidth]{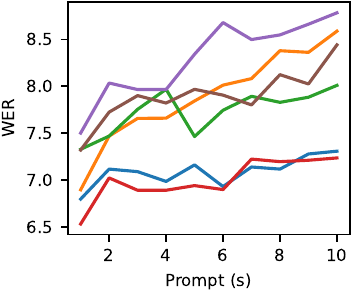}
	\caption{{English}}
	\label{fig:prompt_vs_wer_english}
\end{subfigure}
\hfill
\begin{subfigure}[h]{.32\textwidth}
	\centering
	\includegraphics[width=\textwidth]{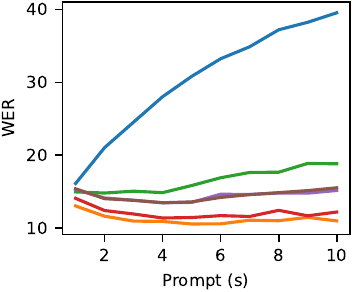}
	\caption{{French}}
	\label{fig:prompt_vs_wer_french}
\end{subfigure}
\hfill
\begin{subfigure}[h]{.32\textwidth}
	\centering
	\includegraphics[width=\textwidth]{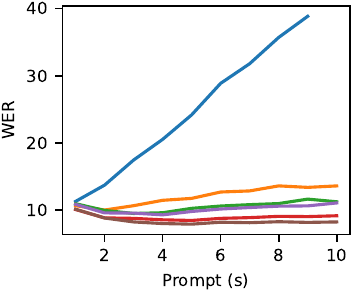}
	\caption{{Spanish}}
	\label{fig:prompt_vs_wer_spanish}
\end{subfigure}
\hfill
\begin{subfigure}[h]{.32\textwidth}
	\centering
	\includegraphics[width=\textwidth]{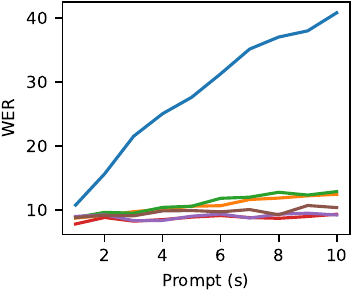}
	\caption{{Portuguese}}
	\label{fig:prompt_vs_wer_portuguese}
\end{subfigure}
\hfill
\begin{subfigure}[h]{.32\textwidth}
	\centering
	\includegraphics[width=\textwidth]{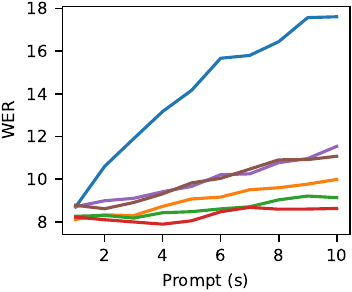}
	\caption{{German}}
	\label{fig:prompt_vs_wer_german}
\end{subfigure}
\hfill
\begin{subfigure}[h]{.32\textwidth}
	\centering
	\includegraphics[width=\textwidth]{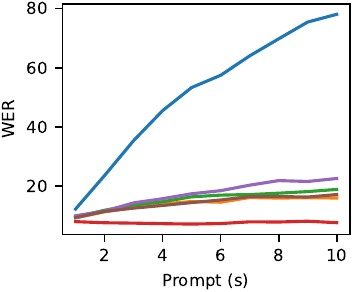}
	\caption{{Polish}}
	\label{fig:prompt_vs_wer_polish}
\end{subfigure}
\caption{Each subplot considers one of the six target language and shows WER as a function of prompt audio duration in seconds for cross-lingual style transfer from different source language. We find WER remain reasonably low for all cases except for ``English'' to ``X'' style transfer.We set the classifier-free guidance strength ($\alpha$) to $1.0$ and use midpoint ODE solver with a NFE of $32$.}
\label{fig:prompt_vs_wer_mls}
\end{figure}

\boldparagraph{Cross-lingual} 
Here we examine the effect of increasing the prompt length for the case of cross-lingual zero-shot TTS. As described in \ref{sec:exp_zs_tts}, this setting has a total 36 language transfer directions for each pair of source and target language. For each target text in a given transfer setting, we examine how WER / SIM-o\footnote{Same trend is observed with SIM-r. We present SIM-o to be consistent with \cref{tab:ttsx}} vary as the prompt length increases. Similarly, the regression duration model is used for the target text. \cref{fig:prompt_vs_similarity_mls} and \cref{fig:prompt_vs_wer_mls} plot the SIM-o (speaker similarity) and WER trends respectively. When concatenating the prompt to the target for MLS, we find that the samples are quite a bit longer than what the model was trained on (16s max length), because MLS test set samples are in average 15 seconds long. To alleviate this out of domain issue and focus the study on varying the prompt length, we truncate the target sequences to 4 seconds (at word boundaries). We notice that WERs are higher compared to \cref{tab:ttsx}, likely because the ASR model struggles with incomplete sentences. Each subplot contains the trend for one of the target languages from all six source languages.

The speaker similarity consistently improves as the prompt length is increased, similar to the monolingual setting.
In contrast, we find that WER increases as we increase the prompt length for most directions. The WER increases much more for En $\rightarrow$ non-En directions. We hypothesize that this is due to training data imbalance across languages, where English accounts for over 90\% of the multilingual training data. Hence, when transferring from English, the model is more likely to assume that the whole sentence is in English as the prompt length increases and produce incorrect pronunciation for the non-English target. Note that during the training phase, the model was only exposed to audio samples and phonemes originating from a single language.

\subsection{Ablation on generative modeling approaches}\label{sec:exp_gen_approach}
We compare three generative modeling approaches in this section: the proposed flow-matching with the OT path (FM w/ OT), flow-matching with the variance preserving (VP) diffusion path (FM w/ diff), and score-matching with the VP diffusion path (SM w/ diff). The last two objectives model the same probability path, but they predict different objects (vector field vs score function). A reduced setup described in~\ref{sec:app_train_obj} with a lowered learning rate (1e-4) and losses on all frames~(\cref{eq:unmasked-loss}) is adopted to ensure convergence for all three objectives. 

We vary the number of training and inference steps, and evaluate models on the zero-shot TTS task(~\cref{sec:exp_zs_tts}). Results in \cref{tab:obj_train_step} shows that FM w/ OT trains significantly faster than the other two objectives, achieving the best performance with 100K training steps, and even outperforms SM w/ diff using only 50K updates. Results in \cref{tab:obj_inf_nfe} shows superior inference efficiency of FM w/ OT, which can produce good results with just 8 NFEs, while FM w/ diff requires at least 16 NFEs and SM w/ diff requires over 64 NFEs. Complete results are in \cref{tab:ablation_method}

\begin{table}[h]
    \centering
    \caption{Comparing different objectives on training efficiency. 32 NFEs are used for inference. Each model is evaluated on the monolingual zero-shot TTS task.}
    \label{tab:obj_train_step}
    \begin{tabular}{l|cccccc}
        \toprule
        \multirow{2}{*}{Method}
        & \multicolumn{2}{c}{upd=50K} 
        & \multicolumn{2}{c}{upd=100K} 
        & \multicolumn{2}{c}{upd=150K} \\
              & WER & Sim-o & WER & Sim-o & WER & Sim-o \\
        \midrule
        FM w/ OT (proposed) & \textbf{2.5}  & \textbf{0.424} & \textbf{2.2}  & \textbf{0.487} & \textbf{2.1} & \textbf{0.508} \\
        FM w/ diff          & 76.0 & 0.066 & 3.1  & 0.344 & 2.6 & 0.478 \\
        SM w/ diff          & 73.3 & 0.062 & 17.4 & 0.176 & 5.1 & 0.349 \\
        \bottomrule
    \end{tabular}
\end{table}

\begin{table}[h]
    \centering
    \caption{Comparing different objectives on inference efficiency. All models are trained for 150K updates. Each model is evaluated on the monolingual zero-shot TTS task.}
    \label{tab:obj_inf_nfe}
    \begin{tabular}{l|cccccccc}
        \toprule
        \multirow{2}{*}{Method}
        & \multicolumn{2}{c}{NFE=4} 
        & \multicolumn{2}{c}{NFE=8} 
        & \multicolumn{2}{c}{NFE=16} 
        & \multicolumn{2}{c}{NFE=32} \\
        & WER & Sim-o & WER & Sim-o & WER & Sim-o & WER & Sim-o \\
        \midrule
        FM w/ OT (proposed) & \textbf{2.4}  & \textbf{0.410} &\textbf{ 2.2}  & \textbf{0.481} & \textbf{2.2}  & \textbf{0.503} & \textbf{2.1} & \textbf{0.508} \\
        FM w/ diff          & 11.5 & 0.171 & 3.0  & 0.359 & 2.7  & 0.447 & 2.6 & 0.478 \\
        SM w/ diff          & 94.5 & 0.054 & 42.3 & 0.076 & 11.5 & 0.218 & 5.1 & 0.349 \\
        \bottomrule
    \end{tabular}
\end{table}

\section{Ethical Statement}\label{sec:fake_detect}
We recognize the potential risks of a model capable of generating speech in the style of arbitrary people.  In an effort to diminish these risks we show that a binary classification model is able to consistently distinguish between real world speech and that which is generated from our model.

Inspired by \citep{spear-tts}, we train a convolutional binary classification model to distinguish between real and generated speech.  The model consists of 6 blocks with hidden dimension sizes: [64, 128, 256, 256, 512, 512].  Each block contains a (3 x 1) convolution along the time axis, a (1 x 3) convolution along the frequency axis, followed by a ReLU activation and batch normalization.  After each block that increases the hidden dimension size we also apply max pooling with a stride of 2 across both the time and frequency dimensions.  Finally, global max pooling is applied and a linear layer projects to a single value that is fed into a binary cross entropy loss.  At inference time we create a sliding window with hop length equal to 250ms and run each chunk of audio through the classifier and average the outputs.  

The model is tested on the dev-clean split of Librispeech. We then take a 100 hour subset of the 60K hour-English data and set aside 2,703 random utterances (to match the size of dev-clean) which is used as a validation split. The remaining utterances from the 100 hours subset are used as the ground truth utterances for training.  For each split we synthesize audio, conditioned on each utterance of the split by masking out frames in the spectrogram corresponding to 90\%, 50\%, and 30\% of the phonemes of the utterance.  All samples are generated using classifier-free guidance with $w = 0.7$, midpoint ODE solver (step size 0.0625 / NFE=64), and the regression duration model. 

We consider two detection tasks. The first one is to distinguish between original audio and \vb{}-generated audio. The second one is to distinguish resynthesized audio and \vb{}-generated audio. The resynthesized audio is created by extracting the Mel Spectrogram from original audio and then vocoding it with the HiFi-GAN vocoder.

\cref{tab:synth-detection} presents the results for each setting. The model can trivially distinguish original audio from \vb{}-generated audio. This results from the fact that a model can also trivially distinguish original audio from resynthesized audio, most likely by recognizing artifacts produced by the vocoder. The task of differentiating \vb{}-generated audio from resynthesized audio is much harder. When 90\% of the audio is masked, the model is able to reliably classify the audio as \vb{}-generated.  In lower masking regimes this decreases a bit, but this is likely due to a naive inference method of averaging the outputs of all sliding windows.  Since the majority of windows are non-synthetic, this leads to mis-classifications.  
\begin{table}[ht]

\caption{Synthetic speech detection metrics}
\label{tab:synth-detection}
\centering

\begin{tabular}{lrrrr}
\toprule
\textbf{\% Mask} &  \textbf{Accuracy} &  \textbf{Precision} &    \textbf{Recall} \\
\midrule
\multicolumn{4}{l}{\textit{Original audio vs \vb{}-generated audio}} \\
   30\% &     1.000 &      1.000 &   1.000 \\
   50\% &     1.000 &      1.000 &   1.000 \\
   90\% &     1.000 &      1.000 &   1.000 \\
\midrule
\multicolumn{4}{l}{\textit{Resynthesized audio vs \vb{}-generated audio}} \\
 30\% &    0.704 &     0.714 &  0.680 \\
 50\% &    0.809 &     0.796 &  0.831 \\
 90\% &    0.907 &     0.881 &  0.942 \\
\bottomrule
\end{tabular}

\end{table}

\section{Conclusion and Discussion}

This paper presents \vb{}, the most versatile generative model for speech. By learning to solve a text-guided speech infilling task on large scale multilingual datasets with a power model and training objective \vb{} demonstrates impressive task generalization capabilities. \vb{} achieves state-of-the-art performance on mono and cross-lingual zero-shot TTS, speech inpainting, and diverse speech sampling, and can generate speech up to 20 times faster than the best autoregressive models.

\paragraph{Limitation}
\vb{} models presented in this paper are trained on read speech from audiobooks in up to six written languages. Hence, the current models may not transfer well to conversational speech~\citep{godfrey1992switchboard}, which is more casual and contains more non-verbal sounds such as laughing and back-channeling (e.g., um-hmm). We plan to tackle the problem by scaling the training data to incorporate more diverse speech.

On the other hand, \vb{} depends on a phonemizer and a forced aligner to produce frame-level phonetic transcript. In addition, many existing phonemizers~\citep{McAuliffe2017MontrealFA} are word-based, which does not take neighboring words of the target into account when predicting the pronunciation. Such phonemizers cannot accurately predict phonetic transcript given text because pronunciation is context-dependent in many languages (e.g., liaisons in French). In the future, we will explore more end-to-end methods where a model would be able to take raw text with punctuation as input~\citep{Casanova2021YourTTSTZ}, and eliminate the need of phonemizers and forced aligners to improve the performance and increase the language coverage.

Last but not least, while \vb{} yields impressive results on transferring audio style (voice, speaking style, emotion, and acoustic condition), the model does not allow independent control of each attribute. In other words, one cannot ask the model to generate speech that resembles voice of one sample while resembling the emotion of another sample. We leave disentangled control of attributes through prompting or text description for future work.

\paragraph{Broader impact} 
A high-quality and versatile generalist speech generation model like \vb{} can enable many applications that improve the quality of our life. For example, zero-shot TTS could bring the voice back to people who suffer from diseases or underwent surgeries such as laryngectomy the causes inability to speak. Zero-shot TTS can also be combined with visual speech recognition systems~\citep{hsu2022revise} to avoid the need of typing. 
When paired with speech translation models, cross-lingual zero-shot TTS enables everyone to speak any language in their own voice. Content editing and speech denoising can be productivity tools for users to create content more effortlessly. Diverse speech sampling, as shown in the paper, can significantly reduces the cost of creating data for training speech-input models.

While \vb{} can bring many positive social impacts, it also carries the potential of misuse and unintended harm. To mitigate the risk, we have presented a highly effective classifier in \cref{sec:fake_detect} showing that the model can accurately distinguish between real and synthetic speech. For future work, we also plan to investigate proactive methods for training the generative model such that the synthetic speech can be more easily detected, such as embedding artificial fingerprints~\citep{yu2021artificial} that can be trivially detected without hurting the speech quality.

To prevent Voicebox from learning biases, we also need to carefully select its training data. First, if Voicebox is only trained on a smaller number of speakers from a specific group with similar accents, it will not be able to generate diverse speech representing the accents around the globe, and downstream models trained on Voicebox generated speech would perform worse on groups with underrepresented accents. For zero-shot style transfer, the performance would also degrade for underrepresented accents. To mitigate this, we have leveraged in-the-wild speech that includes a wide variety of accents, and will continue investing in collecting diverse speech to avoid such biases.

Second, if Voicebox is trained on data where samples from one ethnic group always have lower audio quality (e.g., more noise) while the other ethnic group always has higher audio quality samples, the model would also learn undesired association. To mitigate this, we want the distribution of audio quality (and other audio attributes) and ethnic group to be less correlated, which is usually the case when we have larger scale data collected from in-the-wild sources. We can further tackle this by leveraging data augmentation to decorrelate the distribution, such as adding noise and enhancing speech to widen the audio quality distribution.

\section*{Acknowledgment}
The authors would like to thank Kristin Lauter and Joelle Pineau for supporting the project, 
thank Ricky Chen, Yaron Lipman, Alexandre Defossez, Gabriel Synnaeve for the technical discussion, 
thank Jade Copet and Gabriel Synnaeve for the compute support, 
thank Eleonora Presani and Jackie Pan for discussing the responsible AI studies, 
thank William Ngan, Somya Jain, Lydia Baillergeau, Dana Beaty, Chantal Mora, Daniel Duncan, Gopika Jhala, Steph Miles, Josh Terry, Valeryia Aranovich, Ashton Evans, Aly Gill, Andrea Mileskiewicz, Emily Richards, and Aaron Vasquez for developing the visual assets and the website, 
thank Alyssa Newcomb and Oliver Libaw for developing the posts, 
thank Peter Gray, Natalie Hereth, Shauna Kelleher, Ashley Gabriel, Seine Kim, Ana Paula Kirschner Moffarej and Aiman Farooq for coordinating the launch, 
thank Harrison Rudolph, Mallika Malhotra, Carolyn Krol, Lauren Cohen and Mo Metanat for the reviews, 
and thank 
Alexandra Gualdino,
Ana Paula Kirschner Mofarrej,
Benjamin Muller,
Chloe Rolland,
Daniella Kalfa,
Darcie Da Silva,
Gabriel Synnaeve,
Hunter Goldman,
Juan Pino,
Karen Ulrich,
Kris Sekula,
Manuel Ribeiro,
Marina Zannoli,
Mary Williamson,
Rashel Moritz,
Stephanie Castillo,
Tu Anh Nguyen,
Vlad Sobal,
Volker Seeker,
and anonymous volunteers for sharing speech samples for the demo.

\bibliography{refs}

\begin{thebibliography}{79}
\providecommand{\natexlab}[1]{#1}
\providecommand{\url}[1]{\texttt{#1}}
\expandafter\ifx\csname urlstyle\endcsname\relax
  \providecommand{\doi}[1]{doi: #1}\else
  \providecommand{\doi}{doi: \begingroup \urlstyle{rm}\Url}\fi

\bibitem[Aghajanyan et~al.(2023)Aghajanyan, Yu, Conneau, Hsu, Hambardzumyan,
  Zhang, Roller, Goyal, Levy, and Zettlemoyer]{Aghajanyan2023ScalingLF}
A.~Aghajanyan, L.~Yu, A.~Conneau, W.-N. Hsu, K.~Hambardzumyan, S.~Zhang,
  S.~Roller, N.~Goyal, O.~Levy, and L.~Zettlemoyer.
\newblock Scaling laws for generative mixed-modal language models.
\newblock \emph{ArXiv}, abs/2301.03728, 2023.

\bibitem[Akuzawa et~al.(2018)Akuzawa, Iwasawa, and
  Matsuo]{Akuzawa2018ExpressiveSS}
K.~Akuzawa, Y.~Iwasawa, and Y.~Matsuo.
\newblock Expressive speech synthesis via modeling expressions with variational
  autoencoder.
\newblock \emph{ArXiv}, abs/1804.02135, 2018.

\bibitem[Ardila et~al.(2019)Ardila, Branson, Davis, Henretty, Kohler, Meyer,
  Morais, Saunders, Tyers, and Weber]{Ardila2019CommonVA}
R.~Ardila, M.~Branson, K.~Davis, M.~Henretty, M.~Kohler, J.~Meyer, R.~Morais,
  L.~Saunders, F.~M. Tyers, and G.~Weber.
\newblock Common voice: A massively-multilingual speech corpus.
\newblock In \emph{International Conference on Language Resources and
  Evaluation}, 2019.

\bibitem[Babu et~al.(2022)Babu, Wang, Tjandra, Lakhotia, Xu, Goyal, Singh, von
  Platen, Saraf, Pino, Baevski, Conneau, and Auli]{Babu2022XLSR}
A.~Babu, C.~Wang, A.~Tjandra, K.~Lakhotia, Q.~Xu, N.~Goyal, K.~Singh, P.~von
  Platen, Y.~Saraf, J.~Pino, A.~Baevski, A.~Conneau, and M.~Auli.
\newblock {XLS-R:} self-supervised cross-lingual speech representation learning
  at scale.
\newblock In H.~Ko and J.~H.~L. Hansen, editors, \emph{Interspeech 2022, 23rd
  Annual Conference of the International Speech Communication Association,
  Incheon, Korea, 18-22 September 2022}, pages 2278--2282. {ISCA}, 2022.

\bibitem[Baevski et~al.(2020)Baevski, Zhou, Mohamed, and
  Auli]{baevski2020wav2vec}
A.~Baevski, Y.~Zhou, A.~Mohamed, and M.~Auli.
\newblock wav2vec 2.0: A framework for self-supervised learning of speech
  representations.
\newblock \emph{Advances in neural information processing systems}, 2020.

\bibitem[Bai et~al.(2022)Bai, Zheng, Chen, Li, Ma, and Huang]{Bai2022A3TAA}
H.~Bai, R.~Zheng, J.~Chen, X.~Li, M.~Ma, and L.~Huang.
\newblock {A3T}: Alignment-aware acoustic and text pretraining for speech
  synthesis and editing.
\newblock In \emph{International Conference on Machine Learning}, 2022.

\bibitem[Borsos et~al.(2022{\natexlab{a}})Borsos, Marinier, Vincent,
  Kharitonov, Pietquin, Sharifi, Teboul, Grangier, Tagliasacchi, and
  Zeghidour]{Borsos2022AudioLMAL}
Z.~Borsos, R.~Marinier, D.~Vincent, E.~Kharitonov, O.~Pietquin, M.~Sharifi,
  O.~Teboul, D.~Grangier, M.~Tagliasacchi, and N.~Zeghidour.
\newblock {AudioLM}: a language modeling approach to audio generation.
\newblock \emph{ArXiv}, abs/2209.03143, 2022{\natexlab{a}}.

\bibitem[Borsos et~al.(2022{\natexlab{b}})Borsos, Sharifi, and
  Tagliasacchi]{Borsos2022SpeechPainterTS}
Z.~Borsos, M.~Sharifi, and M.~Tagliasacchi.
\newblock {SpeechPainter}: Text-conditioned speech inpainting.
\newblock In \emph{Interspeech}, 2022{\natexlab{b}}.

\bibitem[Brock et~al.(2018)Brock, Donahue, and Simonyan]{brock2018large}
A.~Brock, J.~Donahue, and K.~Simonyan.
\newblock Large scale gan training for high fidelity natural image synthesis.
\newblock \emph{arXiv preprint arXiv:1809.11096}, 2018.

\bibitem[Brown et~al.(2020)Brown, Mann, Ryder, Subbiah, Kaplan, Dhariwal,
  Neelakantan, Shyam, Sastry, Askell, Agarwal, Herbert-Voss, Krueger, Henighan,
  Child, Ramesh, Ziegler, Wu, Winter, Hesse, Chen, Sigler, Litwin, Gray, Chess,
  Clark, Berner, McCandlish, Radford, Sutskever, and
  Amodei]{Brown2020LanguageMA}
T.~B. Brown, B.~Mann, N.~Ryder, M.~Subbiah, J.~Kaplan, P.~Dhariwal,
  A.~Neelakantan, P.~Shyam, G.~Sastry, A.~Askell, S.~Agarwal, A.~Herbert-Voss,
  G.~Krueger, T.~J. Henighan, R.~Child, A.~Ramesh, D.~M. Ziegler, J.~Wu,
  C.~Winter, C.~Hesse, M.~Chen, E.~Sigler, M.~Litwin, S.~Gray, B.~Chess,
  J.~Clark, C.~Berner, S.~McCandlish, A.~Radford, I.~Sutskever, and D.~Amodei.
\newblock Language models are few-shot learners.
\newblock \emph{ArXiv}, abs/2005.14165, 2020.

\bibitem[Casanova et~al.(2021)Casanova, Weber, Shulby, J{\'u}nior, G{\"o}lge,
  and Ponti]{Casanova2021YourTTSTZ}
E.~Casanova, J.~Weber, C.~D. Shulby, A.~C. J{\'u}nior, E.~G{\"o}lge, and M.~A.
  Ponti.
\newblock {YourTTS}: Towards zero-shot multi-speaker tts and zero-shot voice
  conversion for everyone.
\newblock In \emph{International Conference on Machine Learning}, 2021.

\bibitem[Casanova et~al.(2022)Casanova, Junior, Shulby, Oliveira, Teixeira,
  Ponti, and Alu{\'\i}sio]{casanova2022tts}
E.~Casanova, A.~C. Junior, C.~Shulby, F.~S.~d. Oliveira, J.~P. Teixeira, M.~A.
  Ponti, and S.~Alu{\'\i}sio.
\newblock Tts-portuguese corpus: a corpus for speech synthesis in brazilian
  portuguese.
\newblock \emph{Language Resources and Evaluation}, 56\penalty0 (3):\penalty0
  1043--1055, 2022.

\bibitem[Chen(2018)]{torchdiffeq}
R.~T.~Q. Chen.
\newblock torchdiffeq, 2018.
\newblock URL \url{https://github.com/rtqichen/torchdiffeq}.

\bibitem[Chen et~al.(2018)Chen, Rubanova, Bettencourt, and Duvenaud]{cnf}
R.~T.~Q. Chen, Y.~Rubanova, J.~Bettencourt, and D.~K. Duvenaud.
\newblock Neural ordinary differential equations.
\newblock In \emph{Neural Information Processing Systems}, 2018.

\bibitem[Chen et~al.(2022)Chen, Wang, Chen, Wu, Liu, Chen, Li, Kanda, Yoshioka,
  Xiao, et~al.]{Chen2021WavLMLS}
S.~Chen, C.~Wang, Z.~Chen, Y.~Wu, S.~Liu, Z.~Chen, J.~Li, N.~Kanda,
  T.~Yoshioka, X.~Xiao, et~al.
\newblock Wavlm: Large-scale self-supervised pre-training for full stack speech
  processing.
\newblock \emph{IEEE Journal of Selected Topics in Signal Processing},
  16\penalty0 (6):\penalty0 1505--1518, 2022.

\bibitem[D{\'e}fossez et~al.(2020)D{\'e}fossez, Synnaeve, and
  Adi]{defossez2020real}
A.~D{\'e}fossez, G.~Synnaeve, and Y.~Adi.
\newblock Real time speech enhancement in the waveform domain.
\newblock \emph{ArXiv}, abs/2006.12847, 2020.

\bibitem[D{\'e}fossez et~al.(2022)D{\'e}fossez, Copet, Synnaeve, and
  Adi]{Defossez2022HighFN}
A.~D{\'e}fossez, J.~Copet, G.~Synnaeve, and Y.~Adi.
\newblock High fidelity neural audio compression.
\newblock \emph{ArXiv}, abs/2210.13438, 2022.

\bibitem[Desplanques et~al.(2020)Desplanques, Thienpondt, and
  Demuynck]{desplanques2020ecapa}
B.~Desplanques, J.~Thienpondt, and K.~Demuynck.
\newblock {ECAPA-TDNN: Emphasized Channel Attention, propagation and
  aggregation in TDNN based speaker verification}.
\newblock In \emph{Interspeech}, 2020.

\bibitem[Dhariwal and Nichol(2021)]{dhariwal2021diffusion}
P.~Dhariwal and A.~Nichol.
\newblock Diffusion models beat {GANs} on image synthesis.
\newblock \emph{Advances in Neural Information Processing Systems}, 2021.

\bibitem[Godfrey et~al.(1992)Godfrey, Holliman, and
  McDaniel]{godfrey1992switchboard}
J.~J. Godfrey, E.~C. Holliman, and J.~McDaniel.
\newblock Switchboard: Telephone speech corpus for research and development.
\newblock In \emph{Acoustics, Speech, and Signal Processing, IEEE International
  Conference on}, volume~1, pages 517--520. IEEE Computer Society, 1992.

\bibitem[Gulati et~al.(2020)Gulati, Qin, Chiu, Parmar, Zhang, Yu, Han, Wang,
  Zhang, Wu, et~al.]{gulati2020conformer}
A.~Gulati, J.~Qin, C.-C. Chiu, N.~Parmar, Y.~Zhang, J.~Yu, W.~Han, S.~Wang,
  Z.~Zhang, Y.~Wu, et~al.
\newblock Conformer: Convolution-augmented transformer for speech recognition.
\newblock \emph{arXiv preprint arXiv:2005.08100}, 2020.

\bibitem[Heusel et~al.(2017)Heusel, Ramsauer, Unterthiner, Nessler, and
  Hochreiter]{heusel2017gans}
M.~Heusel, H.~Ramsauer, T.~Unterthiner, B.~Nessler, and S.~Hochreiter.
\newblock {GANs} trained by a two time-scale update rule converge to a local
  {Nash} equilibrium.
\newblock \emph{Advances in neural information processing systems}, 2017.

\bibitem[Ho and Salimans(2022)]{ho2022classifier}
J.~Ho and T.~Salimans.
\newblock Classifier-free diffusion guidance.
\newblock \emph{arXiv preprint arXiv:2207.12598}, 2022.

\bibitem[Ho et~al.(2020)Ho, Jain, and Abbeel]{Ho2020DenoisingDP}
J.~Ho, A.~Jain, and P.~Abbeel.
\newblock Denoising diffusion probabilistic models.
\newblock \emph{Advances in Neural Information Processing Systems}, 2020.

\bibitem[Hoffmann et~al.(2022)Hoffmann, Borgeaud, Mensch, Buchatskaya, Cai,
  Rutherford, de~Las~Casas, Hendricks, Welbl, Clark, Hennigan, Noland,
  Millican, van~den Driessche, Damoc, Guy, Osindero, Simonyan, Elsen, Rae,
  Vinyals, and Sifre]{Hoffmann2022TrainingCL}
J.~Hoffmann, S.~Borgeaud, A.~Mensch, E.~Buchatskaya, T.~Cai, E.~Rutherford,
  D.~de~Las~Casas, L.~A. Hendricks, J.~Welbl, A.~Clark, T.~Hennigan, E.~Noland,
  K.~Millican, G.~van~den Driessche, B.~Damoc, A.~Guy, S.~Osindero,
  K.~Simonyan, E.~Elsen, J.~W. Rae, O.~Vinyals, and L.~Sifre.
\newblock Training compute-optimal large language models.
\newblock \emph{ArXiv}, abs/2203.15556, 2022.

\bibitem[Hsu et~al.(2019)Hsu, Zhang, Weiss, Zen, Wu, Wang, Cao, Jia, Chen,
  Shen, et~al.]{Hsu2018HierarchicalGM}
W.-N. Hsu, Y.~Zhang, R.~J. Weiss, H.~Zen, Y.~Wu, Y.~Wang, Y.~Cao, Y.~Jia,
  Z.~Chen, J.~Shen, et~al.
\newblock Hierarchical generative modeling for controllable speech synthesis.
\newblock In \emph{International Conference on Learning Representations}, 2019.

\bibitem[Hsu et~al.(2021)Hsu, Bolte, Tsai, Lakhotia, Salakhutdinov, and
  Mohamed]{Hsu2021HuBERTSS}
W.-N. Hsu, B.~Bolte, Y.-H.~H. Tsai, K.~Lakhotia, R.~Salakhutdinov, and
  A.~Mohamed.
\newblock Hubert: Self-supervised speech representation learning by masked
  prediction of hidden units.
\newblock \emph{IEEE/ACM Transactions on Audio, Speech, and Language
  Processing}, 29:\penalty0 3451--3460, 2021.

\bibitem[Hsu et~al.(2022)Hsu, Remez, Shi, Donley, and Adi]{hsu2022revise}
W.-N. Hsu, T.~Remez, B.~Shi, J.~Donley, and Y.~Adi.
\newblock Revise: Self-supervised speech resynthesis with visual input for
  universal and generalized speech enhancement.
\newblock \emph{arXiv preprint arXiv:2212.11377}, 2022.

\bibitem[Huang et~al.(2022)Huang, Lam, Wang, Su, Yu, Ren, and
  Zhao]{Huang2022FastDiffAF}
R.~Huang, M.~W.~Y. Lam, J.~Wang, D.~Su, D.~Yu, Y.~Ren, and Z.~Zhao.
\newblock {FastDiff}: A fast conditional diffusion model for high-quality
  speech synthesis.
\newblock In \emph{International Joint Conference on Artificial Intelligence},
  2022.

\bibitem[Jia et~al.(2018)Jia, Zhang, Weiss, Wang, Shen, Ren, Nguyen, Pang,
  Lopez~Moreno, Wu, et~al.]{Jia2018TransferLF}
Y.~Jia, Y.~Zhang, R.~Weiss, Q.~Wang, J.~Shen, F.~Ren, P.~Nguyen, R.~Pang,
  I.~Lopez~Moreno, Y.~Wu, et~al.
\newblock Transfer learning from speaker verification to multispeaker
  text-to-speech synthesis.
\newblock \emph{Advances in neural information processing systems}, 2018.

\bibitem[Kahn et~al.(2019)Kahn, Rivi{\`e}re, Zheng, Kharitonov, Xu, Mazar'e,
  Karadayi, Liptchinsky, Collobert, Fuegen, Likhomanenko, Synnaeve, Joulin,
  rahman Mohamed, and Dupoux]{Kahn2019LibriLightAB}
J.~Kahn, M.~Rivi{\`e}re, W.~Zheng, E.~Kharitonov, Q.~Xu, P.-E. Mazar'e,
  J.~Karadayi, V.~Liptchinsky, R.~Collobert, C.~Fuegen, T.~Likhomanenko,
  G.~Synnaeve, A.~Joulin, A.~rahman Mohamed, and E.~Dupoux.
\newblock {Libri-Light}: A benchmark for asr with limited or no supervision.
\newblock \emph{International Conference on Acoustics, Speech and Signal
  Processing}, 2019.

\bibitem[Kameoka et~al.(2018)Kameoka, Kaneko, Tanaka, and
  Hojo]{Kameoka2018StarGANVCNM}
H.~Kameoka, T.~Kaneko, K.~Tanaka, and N.~Hojo.
\newblock {StarGAN-VC}: non-parallel many-to-many voice conversion using star
  generative adversarial networks.
\newblock \emph{IEEE Spoken Language Technology Workshop}, 2018.

\bibitem[Kharitonov et~al.(2021)Kharitonov, Lee, Polyak, Adi, Copet, Lakhotia,
  Nguyen, Rivi{\`e}re, rahman Mohamed, Dupoux, and
  Hsu]{Kharitonov2021TextFreePG}
E.~Kharitonov, A.~Lee, A.~Polyak, Y.~Adi, J.~Copet, K.~Lakhotia, T.~Nguyen,
  M.~Rivi{\`e}re, A.~rahman Mohamed, E.~Dupoux, and W.-N. Hsu.
\newblock Text-free prosody-aware generative spoken language modeling.
\newblock In \emph{Annual Meeting of the Association for Computational
  Linguistics}, 2021.

\bibitem[Kharitonov et~al.(2023)Kharitonov, Vincent, Borsos, Marinier, Girgin,
  Pietquin, Sharifi, Tagliasacchi, and Zeghidour]{spear-tts}
E.~Kharitonov, D.~Vincent, Z.~Borsos, R.~Marinier, S.~Girgin, O.~Pietquin,
  M.~Sharifi, M.~Tagliasacchi, and N.~Zeghidour.
\newblock Speak, read and prompt: High-fidelity text-to-speech with minimal
  supervision, 2023.

\bibitem[Kilgour et~al.(2019)Kilgour, Zuluaga, Roblek, and
  Sharifi]{Kilgour2019FrchetAD}
K.~Kilgour, M.~Zuluaga, D.~Roblek, and M.~Sharifi.
\newblock Fr{\'e}chet audio distance: A reference-free metric for evaluating
  music enhancement algorithms.
\newblock In \emph{Interspeech}, 2019.

\bibitem[Kim et~al.(2020)Kim, Kim, Kong, and Yoon]{Kim2020GlowTTSAG}
J.~Kim, S.~Kim, J.~Kong, and S.~Yoon.
\newblock {Glow-TTS}: A generative flow for text-to-speech via monotonic
  alignment search.
\newblock \emph{Advances in Neural Information Processing Systems}, 2020.

\bibitem[Kim et~al.(2021)Kim, Kong, and Son]{Kim2021ConditionalVA}
J.~Kim, J.~Kong, and J.~Son.
\newblock Conditional variational autoencoder with adversarial learning for
  end-to-end text-to-speech.
\newblock In \emph{International Conference on Machine Learning}, 2021.

\bibitem[Kingma and Ba(2014)]{Kingma2014AdamAM}
D.~P. Kingma and J.~Ba.
\newblock Adam: A method for stochastic optimization.
\newblock \emph{CoRR}, abs/1412.6980, 2014.

\bibitem[Kingma and Dhariwal(2018)]{kingma2018glow}
D.~P. Kingma and P.~Dhariwal.
\newblock Glow: Generative flow with invertible 1x1 convolutions.
\newblock \emph{Advances in neural information processing systems}, 31, 2018.

\bibitem[Kreuk et~al.(2022)Kreuk, Polyak, Copet, Kharitonov, Nguyen,
  Rivi{\`e}re, Hsu, Mohamed, Dupoux, and Adi]{kreuk2021textless}
F.~Kreuk, A.~Polyak, J.~Copet, E.~Kharitonov, T.-A. Nguyen, M.~Rivi{\`e}re,
  W.-N. Hsu, A.~Mohamed, E.~Dupoux, and Y.~Adi.
\newblock Textless speech emotion conversion using decomposed and discrete
  representations.
\newblock In \emph{Proceedings of the 2022 Conference on Empirical Methods in
  Natural Language Processing}, 2022.

\bibitem[Kubichek(1993)]{kubichek1993mel}
R.~Kubichek.
\newblock Mel-cepstral distance measure for objective speech quality
  assessment.
\newblock In \emph{Proceedings of IEEE pacific rim conference on communications
  computers and signal processing}, volume~1, pages 125--128. IEEE, 1993.

\bibitem[Lakhotia et~al.(2021)Lakhotia, Kharitonov, Hsu, Adi, Polyak, Bolte,
  Nguyen, Copet, Baevski, Mohamed, and Dupoux]{Lakhotia2021OnGS}
K.~Lakhotia, E.~Kharitonov, W.-N. Hsu, Y.~Adi, A.~Polyak, B.~Bolte, T.~Nguyen,
  J.~Copet, A.~Baevski, A.~B. Mohamed, and E.~Dupoux.
\newblock On generative spoken language modeling from raw audio.
\newblock \emph{Transactions of the Association for Computational Linguistics},
  9:\penalty0 1336--1354, 2021.

\bibitem[{\L}a{\'n}cucki(2021)]{fastpitch}
A.~{\L}a{\'n}cucki.
\newblock Fastpitch: Parallel text-to-speech with pitch prediction.
\newblock In \emph{International Conference on Acoustics, Speech and Signal
  Processing}, 2021.

\bibitem[Le~Roux et~al.(2019)Le~Roux, Wisdom, Erdogan, and Hershey]{le2019sdr}
J.~Le~Roux, S.~Wisdom, H.~Erdogan, and J.~R. Hershey.
\newblock Sdr--half-baked or well done?
\newblock In \emph{ICASSP 2019-2019 IEEE International Conference on Acoustics,
  Speech and Signal Processing (ICASSP)}, pages 626--630. IEEE, 2019.

\bibitem[Lipman et~al.(2023)Lipman, Chen, Ben-Hamu, Nickel, and
  Le]{flow-matching}
Y.~Lipman, R.~T.~Q. Chen, H.~Ben-Hamu, M.~Nickel, and M.~Le.
\newblock Flow matching for generative modeling.
\newblock In \emph{International Conference on Learning Representations}, 2023.

\bibitem[Lorenzo-Trueba et~al.(2018)Lorenzo-Trueba, Yamagishi, Toda, Saito,
  Villavicencio, Kinnunen, and Ling]{LorenzoTrueba2018TheVC}
J.~Lorenzo-Trueba, J.~Yamagishi, T.~Toda, D.~Saito, F.~Villavicencio, T.~H.
  Kinnunen, and Z.~Ling.
\newblock The voice conversion challenge 2018: Promoting development of
  parallel and nonparallel methods.
\newblock \emph{ArXiv}, abs/1804.04262, 2018.

\bibitem[McAuliffe et~al.(2017)McAuliffe, Socolof, Mihuc, Wagner, and
  Sonderegger]{McAuliffe2017MontrealFA}
M.~McAuliffe, M.~Socolof, S.~Mihuc, M.~Wagner, and M.~Sonderegger.
\newblock Montreal forced aligner: Trainable text-speech alignment using kaldi.
\newblock In \emph{Interspeech}, 2017.

\bibitem[Nguyen et~al.(2022)Nguyen, Kharitonov, Copet, Adi, Hsu, Elkahky,
  Tomasello, Algayres, Sagot, Mohamed, and Dupoux]{Nguyen2022GenerativeSD}
T.~Nguyen, E.~Kharitonov, J.~Copet, Y.~Adi, W.-N. Hsu, A.~M. Elkahky,
  P.~Tomasello, R.~Algayres, B.~Sagot, A.~Mohamed, and E.~Dupoux.
\newblock Generative spoken dialogue language modeling.
\newblock \emph{Transactions of the Association for Computational Linguistics},
  11:\penalty0 250--266, 2022.

\bibitem[Nichol et~al.(2021)Nichol, Dhariwal, Ramesh, Shyam, Mishkin, McGrew,
  Sutskever, and Chen]{Nichol2021GLIDETP}
A.~Nichol, P.~Dhariwal, A.~Ramesh, P.~Shyam, P.~Mishkin, B.~McGrew,
  I.~Sutskever, and M.~Chen.
\newblock {GLIDE}: Towards photorealistic image generation and editing with
  text-guided diffusion models.
\newblock In \emph{International Conference on Machine Learning}, 2021.

\bibitem[Panayotov et~al.(2015)Panayotov, Chen, Povey, and
  Khudanpur]{Panayotov2015LibrispeechAA}
V.~Panayotov, G.~Chen, D.~Povey, and S.~Khudanpur.
\newblock Librispeech: An asr corpus based on public domain audio books.
\newblock \emph{International Conference on Acoustics, Speech and Signal
  Processing}, 2015.

\bibitem[Park et~al.(2019)Park, Chan, Zhang, Chiu, Zoph, Cubuk, and
  Le]{park2019specAugmentAS}
D.~S. Park, W.~Chan, Y.~Zhang, C.-C. Chiu, B.~Zoph, E.~D. Cubuk, and Q.~V. Le.
\newblock {SpecAugment}: A simple data augmentation method for automatic speech
  recognition.
\newblock In \emph{Interspeech}, 2019.

\bibitem[Paszke et~al.(2019)Paszke, Gross, Massa, Lerer, Bradbury, Chanan,
  Killeen, Lin, Gimelshein, Antiga, et~al.]{paszke2019pytorch}
A.~Paszke, S.~Gross, F.~Massa, A.~Lerer, J.~Bradbury, G.~Chanan, T.~Killeen,
  Z.~Lin, N.~Gimelshein, L.~Antiga, et~al.
\newblock {Pytorch}: An imperative style, high-performance deep learning
  library.
\newblock \emph{Advances in neural information processing systems}, 2019.

\bibitem[Polyak et~al.(2021)Polyak, Adi, Copet, Kharitonov, Lakhotia, Hsu,
  Mohamed, and Dupoux]{polyak2021speech}
A.~Polyak, Y.~Adi, J.~Copet, E.~Kharitonov, K.~Lakhotia, W.-N. Hsu, A.~Mohamed,
  and E.~Dupoux.
\newblock Speech resynthesis from discrete disentangled self-supervised
  representations.
\newblock In \emph{Interspeech}, 2021.

\bibitem[Popov et~al.(2021)Popov, Vovk, Gogoryan, Sadekova, and
  Kudinov]{Popov2021GradTTSAD}
V.~Popov, I.~Vovk, V.~Gogoryan, T.~Sadekova, and M.~Kudinov.
\newblock {Grad-TTS}: A diffusion probabilistic model for text-to-speech.
\newblock In \emph{International Conference on Machine Learning}, 2021.

\bibitem[Povey et~al.(2011)Povey, Ghoshal, Boulianne, Burget, Glembek, Goel,
  Hannemann, Motlicek, Qian, Schwarz, et~al.]{povey2011kaldi}
D.~Povey, A.~Ghoshal, G.~Boulianne, L.~Burget, O.~Glembek, N.~Goel,
  M.~Hannemann, P.~Motlicek, Y.~Qian, P.~Schwarz, et~al.
\newblock The kaldi speech recognition toolkit.
\newblock In \emph{Workshop on automatic speech recognition and understanding},
  2011.

\bibitem[Press et~al.(2021)Press, Smith, and Lewis]{Press2021TrainST}
O.~Press, N.~A. Smith, and M.~Lewis.
\newblock Train short, test long: Attention with linear biases enables input
  length extrapolation.
\newblock \emph{ArXiv}, abs/2108.12409, 2021.

\bibitem[Radford et~al.(2022)Radford, Kim, Xu, Brockman, McLeavey, and
  Sutskever]{Radford2022RobustSR}
A.~Radford, J.~W. Kim, T.~Xu, G.~Brockman, C.~McLeavey, and I.~Sutskever.
\newblock Robust speech recognition via large-scale weak supervision.
\newblock \emph{ArXiv}, abs/2212.04356, 2022.

\bibitem[Ramesh et~al.(2021)Ramesh, Pavlov, Goh, Gray, Voss, Radford, Chen, and
  Sutskever]{Ramesh2021ZeroShotTG}
A.~Ramesh, M.~Pavlov, G.~Goh, S.~Gray, C.~Voss, A.~Radford, M.~Chen, and
  I.~Sutskever.
\newblock Zero-shot text-to-image generation.
\newblock \emph{ArXiv}, abs/2102.12092, 2021.

\bibitem[Ren et~al.(2021)Ren, Hu, Tan, Qin, Zhao, Zhao, and
  Liu]{Ren2020FastSpeech2F}
Y.~Ren, C.~Hu, X.~Tan, T.~Qin, S.~Zhao, Z.~Zhao, and T.-Y. Liu.
\newblock Fastspeech 2: Fast and high-quality end-to-end text to speech.
\newblock In \emph{International Conference on Learning Representations}, 2021.

\bibitem[Ribeiro et~al.(2011)Ribeiro, Flor{\^e}ncio, Zhang, and
  Seltzer]{ribeiro2011crowdmos}
F.~Ribeiro, D.~Flor{\^e}ncio, C.~Zhang, and M.~Seltzer.
\newblock {CrowdMOS}: An approach for crowdsourcing mean opinion score studies.
\newblock In \emph{International Conference on Acoustics, Speech and Signal
  Processing}, 2011.

\bibitem[Robinson et~al.(2019)Robinson, Obin, and Roebel]{robinson2019sequence}
C.~Robinson, N.~Obin, and A.~Roebel.
\newblock Sequence-to-sequence modelling of {F0} for speech emotion conversion.
\newblock In \emph{International Conference on Acoustics, Speech and Signal
  Processing}, 2019.

\bibitem[Rombach et~al.(2022)Rombach, Blattmann, Lorenz, Esser, and
  Ommer]{rombach2022high}
R.~Rombach, A.~Blattmann, D.~Lorenz, P.~Esser, and B.~Ommer.
\newblock High-resolution image synthesis with latent diffusion models.
\newblock In \emph{Proceedings of the IEEE/CVF Conference on Computer Vision
  and Pattern Recognition}, 2022.

\bibitem[Saharia et~al.(2022)Saharia, Chan, Chang, Lee, Ho, Salimans, Fleet,
  and Norouzi]{saharia2022palette}
C.~Saharia, W.~Chan, H.~Chang, C.~Lee, J.~Ho, T.~Salimans, D.~Fleet, and
  M.~Norouzi.
\newblock Palette: Image-to-image diffusion models.
\newblock In \emph{ACM SIGGRAPH 2022 Conference Proceedings}, 2022.

\bibitem[Serr{\`a} et~al.(2022)Serr{\`a}, Pascual, Pons, Araz, and
  Scaini]{Serr2022UniversalSE}
J.~Serr{\`a}, S.~Pascual, J.~Pons, R.~O. Araz, and D.~Scaini.
\newblock Universal speech enhancement with score-based diffusion.
\newblock \emph{ArXiv}, abs/2206.03065, 2022.

\bibitem[Shen et~al.(2017)Shen, Pang, Weiss, Schuster, Jaitly, Yang, Chen,
  Zhang, Wang, Skerry-Ryan, Saurous, Agiomyrgiannakis, and
  Wu]{Shen2017NaturalTS}
J.~Shen, R.~Pang, R.~J. Weiss, M.~Schuster, N.~Jaitly, Z.~Yang, Z.~Chen,
  Y.~Zhang, Y.~Wang, R.~J. Skerry-Ryan, R.~A. Saurous, Y.~Agiomyrgiannakis, and
  Y.~Wu.
\newblock Natural {TTS} synthesis by conditioning wavenet on mel spectrogram
  predictions.
\newblock \emph{International Conference on Acoustics, Speech and Signal
  Processing}, 2017.

\bibitem[Shen et~al.(2023)Shen, Ju, Tan, Liu, Leng, He, Qin, Zhao, and
  Bian]{shen2023naturalspeech}
K.~Shen, Z.~Ju, X.~Tan, Y.~Liu, Y.~Leng, L.~He, T.~Qin, S.~Zhao, and J.~Bian.
\newblock Naturalspeech 2: Latent diffusion models are natural and zero-shot
  speech and singing synthesizers.
\newblock \emph{arXiv preprint arXiv:2304.09116}, 2023.

\bibitem[Skerry-Ryan et~al.(2018)Skerry-Ryan, Battenberg, Xiao, Wang, Stanton,
  Shor, Weiss, Clark, and Saurous]{skerry2018towards}
R.~Skerry-Ryan, E.~Battenberg, Y.~Xiao, Y.~Wang, D.~Stanton, J.~Shor, R.~Weiss,
  R.~Clark, and R.~A. Saurous.
\newblock Towards end-to-end prosody transfer for expressive speech synthesis
  with tacotron.
\newblock In \emph{international conference on machine learning}, pages
  4693--4702. PMLR, 2018.

\bibitem[Song and Ermon(2019)]{song2019generative}
Y.~Song and S.~Ermon.
\newblock Generative modeling by estimating gradients of the data distribution.
\newblock \emph{Advances in neural information processing systems}, 32, 2019.

\bibitem[Tan et~al.(2022)Tan, Chen, Liu, Cong, Zhang, Liu, Wang, Leng, Yi, He,
  Soong, Qin, Zhao, and Liu]{Tan2022NaturalSpeechET}
X.~Tan, J.~Chen, H.~Liu, J.~Cong, C.~Zhang, Y.~Liu, X.~Wang, Y.~Leng, Y.~Yi,
  L.~He, F.~K. Soong, T.~Qin, S.~Zhao, and T.-Y. Liu.
\newblock {NaturalSpeech}: End-to-end text to speech synthesis with human-level
  quality.
\newblock \emph{ArXiv}, abs/2205.04421, 2022.

\bibitem[Vaswani et~al.(2017)Vaswani, Shazeer, Parmar, Uszkoreit, Jones, Gomez,
  Kaiser, and Polosukhin]{Vaswani2017AttentionIA}
A.~Vaswani, N.~M. Shazeer, N.~Parmar, J.~Uszkoreit, L.~Jones, A.~N. Gomez,
  L.~Kaiser, and I.~Polosukhin.
\newblock Attention is all you need.
\newblock \emph{ArXiv}, abs/1706.03762, 2017.

\bibitem[Wang et~al.(2021)Wang, Hsu, Adi, Polyak, Lee, Chen, Gu, and
  Pino]{Wang2021fairseqSA}
C.~Wang, W.-N. Hsu, Y.~Adi, A.~Polyak, A.~Lee, P.-J. Chen, J.~Gu, and J.~M.
  Pino.
\newblock fairseq s$^2$: A scalable and integrable speech synthesis toolkit.
\newblock In \emph{Conference on Empirical Methods in Natural Language
  Processing}, 2021.

\bibitem[Wang et~al.(2023)Wang, Chen, Wu, Zhang, Zhou, Liu, Chen, Liu, Wang,
  Li, He, Zhao, and Wei]{Wang2023NeuralCL}
C.~Wang, S.~Chen, Y.~Wu, Z.-H. Zhang, L.~Zhou, S.~Liu, Z.~Chen, Y.~Liu,
  H.~Wang, J.~Li, L.~He, S.~Zhao, and F.~Wei.
\newblock Neural codec language models are zero-shot text to speech
  synthesizers.
\newblock \emph{ArXiv}, abs/2301.02111, 2023.

\bibitem[Wang et~al.(2018)Wang, Stanton, Zhang, Skerry-Ryan, Battenberg, Shor,
  Xiao, Ren, Jia, and Saurous]{Wang2018StyleTU}
Y.~Wang, D.~Stanton, Y.~Zhang, R.~J. Skerry-Ryan, E.~Battenberg, J.~Shor,
  Y.~Xiao, F.~Ren, Y.~Jia, and R.~A. Saurous.
\newblock Style tokens: Unsupervised style modeling, control and transfer in
  end-to-end speech synthesis.
\newblock In \emph{International Conference on Machine Learning}, 2018.

\bibitem[Xu et~al.(2014)Xu, Du, Dai, and Lee]{xu2014regression}
Y.~Xu, J.~Du, L.-R. Dai, and C.-H. Lee.
\newblock A regression approach to speech enhancement based on deep neural
  networks.
\newblock \emph{IEEE/ACM Transactions on Audio, Speech, and Language
  Processing}, 23\penalty0 (1):\penalty0 7--19, 2014.

\bibitem[Yamagishi et~al.(2019)Yamagishi, Veaux, and
  MacDonald]{Yamagishi2019CSTRVC}
J.~Yamagishi, C.~Veaux, and K.~MacDonald.
\newblock Cstr vctk corpus: English multi-speaker corpus for cstr voice cloning
  toolkit (version 0.92).
\newblock 2019.

\bibitem[Yamamoto et~al.(2020)Yamamoto, Song, and Kim]{yamamoto2020parallel}
R.~Yamamoto, E.~Song, and J.-M. Kim.
\newblock {Parallel WaveGAN}: A fast waveform generation model based on
  generative adversarial networks with multi-resolution spectrogram.
\newblock In \emph{International Conference on Acoustics, Speech and Signal
  Processing}, 2020.

\bibitem[Yu et~al.(2021)Yu, Skripniuk, Abdelnabi, and Fritz]{yu2021artificial}
N.~Yu, V.~Skripniuk, S.~Abdelnabi, and M.~Fritz.
\newblock Artificial fingerprinting for generative models: Rooting deepfake
  attribution in training data.
\newblock In \emph{Proceedings of the IEEE/CVF International conference on
  computer vision}, pages 14448--14457, 2021.

\bibitem[Zeghidour et~al.(2022)Zeghidour, Luebs, Omran, Skoglund, and
  Tagliasacchi]{Zeghidour2022SoundStreamAE}
N.~Zeghidour, A.~Luebs, A.~Omran, J.~Skoglund, and M.~Tagliasacchi.
\newblock Soundstream: An end-to-end neural audio codec.
\newblock \emph{IEEE/ACM Transactions on Audio, Speech, and Language
  Processing}, 30:\penalty0 495--507, 2022.

\bibitem[Zen et~al.(2019)Zen, Dang, Clark, Zhang, Weiss, Jia, Chen, and
  Wu]{zen2019libritts}
H.~Zen, V.~Dang, R.~Clark, Y.~Zhang, R.~J. Weiss, Y.~Jia, Z.~Chen, and Y.~Wu.
\newblock Libritts: A corpus derived from librispeech for text-to-speech.
\newblock \emph{arXiv preprint arXiv:1904.02882}, 2019.

\end{thebibliography}
\bibliographystyle{abbrvnat}


\clearpage
\appendix

\setcounter{table}{0}
\renewcommand{\thetable}{\Alph{section}\arabic{table}}
\setcounter{figure}{0}
\renewcommand{\thefigure}{\Alph{section}\arabic{figure}}

\section{Additional Details of Experiment Setup}

\subsection{Vocoder}\label{sec:app_vocoder}
We adapt the HiFi-GAN V1 configuration to generate 16kHz audio from 80 dimensional log Mel spectral features sampled at 100Hz. 
To compute the log Mel spectrogram, we use a 1024-point short time Fourier transform with a 640-sample (40ms) analysis window, 160-sample (10ms) shift, and the Hann windowing function to compute the amplitude spectrogram, and then apply an 80 dimension Mel filter with a cutoff frequency at 8kHz. The original HiFi-GAN V1 has four transposed convolution blocks for upsampling. The upsampling factors are $[8,8,2,2]$ and the corresponding kernel sizes are $[16,16,4,4]$. Here we only need a total upsampling factor of 160 instead of 256, and we adjust the upsampling factors to $[5,4,4,2]$ and kernel sizes to $[11,8,8,4]$ accordingly. The other parameters are identical to the HiFi-GAN V1 configuration. Total number of parameters is 13M. We train the adapted HiFi-GAN on the 60K hours of English audiobook data for 1.5M updates on 8 GPUs, which takes 7.5 days.

\subsection{Phone representation}\label{sec:app_phone_repr}
\paragraph{Ghost silence} The frame-level phonetic transcript used for training is obtained through force-aligning speech and phonetic transcript. In particular, a forced aligner may align some frames to a special phone ``\texttt{SIL}'' for non-speech frames (silence or noise). For most forced aligners, only frames between words and frames at the beginning and at the end of an utterance can be aligned to \texttt{SIL}.

During inference, we are only given the text transcript, which does not tell us where we should insert silence to. Hence, it is desired to have the duration model not only predict the duration for each phone (\texttt{SIL} included), but also predict the \textit{existence} of \texttt{SIL} at eligible locations (between words and at the two ends of the utterance).
To tackle it, we introduce \textit{ghost silence} to our phonetic transcript, which are silences in between words with duration of zero frames.

To give an example, suppose the transcript contains three words: ``\texttt{Hey what's up}'' with pronunciation ``\texttt{\{Hey:[A,B], what's:[C], up:[D,E,F]\}}'', and the frame-level phonetic transcript $z$ obtained through forced alignment is $z=(\text{\texttt{SIL A B B SIL C D D D E E F SIL SIL}})$. The phonetic transcripts becomes $y=(\text{\texttt{SIL A B SIL C \textcolor{green}{SIL} D E F SIL}})$, where the ghost silence is highlighted in green. The corresponding duration would be $l=(1, 1, 2, 1, 1, \textcolor{green}{0}, 3, 2, 1, 2)$. A ghost silence is inserted between \texttt{what's} and \texttt{up} during training, and the duration model should predict the duration of it as zero to indicate that there should not be a pause between the two words.

\paragraph{Word-position-dependent phone} The possible absence of silence between words in the frame-level phone transcript can make it hard for the audio model to identify word boundaries.
To help the audio model identify the word boundary which is important when reading a sentence, we introduce word-position-dependent phones which are commonly used in Hidden Markov Model based acoustic models for speech recognition~\citep{povey2011kaldi}. This adds a postfix to each phone in the transcript to denote where it is in the corresponding word. There are four postfixes: \texttt{\_B} for beginning, \texttt{\_E} for end, \texttt{\_I} for intermediate, and \texttt{\_S} for singleton. The above example becomes ``\texttt{\{Hey:[A\_B,B\_E], what's:[C\_S], up:[D\_B,E\_I,F\_E]\}}'' with frame-level phonetic transcript $z=(\text{\texttt{SIL A\_B B\_E B\_E SIL C\_S D\_B D\_B D\_B E\_I E\_I F\_E SIL SIL}})$.

\paragraph{Phone-level mask} In terms of masking, given duration $l$, the relationship of phone-level mask $m'$ and frame-level mask $m$ can be written as $m = \text{\texttt{rep}}(m', l)$. For the applications where a duration model is involved (zero-shot TTS, content editing, diverse speech sampling), the frame-level mask $m$ is extended such that no phone is partially masked. In other words, all the frames corresponding to a phone is either entirely masked or entirely unmasked.
During training, we mask a contiguous chunk of audio, infilling of which is a more challenging task compared to infilling multiple smaller segments. All frames that are aligned to a phone are either entirely masked or unmasked.  Note that masking all frames for a phone is not a necessity but was chosen due to ease of implementation.

\subsection{Data transformation}\label{sec:app_data_transform}
The Mel spectrogram is normalized with the global mean (-5.8843) and standard deviation (2.2615) to stabilize training. The statistics are estimated on 30k random training samples from the 1K hours of English audio.
Input and output duration are dequantized ($x \sim \mathcal{U}[x-0.5,x+0.5]$) and transformed with $\log(1+x)$ following~\citep{Ren2020FastSpeech2F}. Prediction of duration is quantized and clipped such that the minimal duration is greater than or equal to zero.

\subsection{Cross-lingual zero-shot TTS test data filtering}\label{sec:app_mls_filter}
We create a test set for each language by selecting samples from the MLS test split which have Whisper transcription WER lower than 20\% (or 30\% for Polish and Portugueses test splits which contains less than 1K samples), because we found MLS test set contains many examples with incomplete transcriptions missing a large portion of the utterance. In addition, a small amount of utterances were excluded due to MFA alignment failure. \cref{tab:mls_num_test} lists the number of samples remained for each language.

\begin{table}[h]
    \centering
    \caption{Number of MLS test samples after filtering.}\label{tab:mls_num_test}
    \begin{tabular}{c|cc}
    \toprule
    \textbf{Language} & \textbf{\#samples before filtering} & \textbf{\#samples after filtering} \\
    \midrule
    English & 3769 & 3535 \\
    Spanish & 2385 & 2323 \\
    German & 3394 & 3183 \\
    French & 2426 & 2284 \\
    Polish & 520 & 508 \\
    Portuguese & 871 & 838 \\
    \bottomrule
    \end{tabular}
\end{table}

\subsection{Setup for training ASR models with synthetic speech}
\label{sec:app_asr_details}

To train an ASR model in \cref{sec:exp_speech_samp}, we extract 80-dimensional log Mel features with a 25ms window and a 10ms frame shift, and then apply global mean-variance normalization. The ASR model is an RNN-T with a Conformer-based encoder~\citep{gulati2020conformer}. The conformer applies time scale reduction to the input features with stride 6, embeds them into 512-dimensional vectors, passes these vectors through a 20-layer conformer which has 8 attention heads and 2048-dimensional fully-connected layers. The conformer output is further mapped to 1024 dimensions through a linear layer followed by layer normalization before being passed to the joiner. The predictor of the network first embeds wordpiece units into 512 dimensional embeddings, applies layer normalization, a 512-dimensional LSTM, a dropout layer and a linear layer that maps the LSTM output to 1024 dimensions. The joiner adds the encoder and predictor outputs, applies tanh non-linearity and uses a linear layer that maps the 512-dimensional joiner input into wordpiece units. There are 4096 wordpiece units estimated from the LibriSpeech 960hr training text.

We apply SpecAugment~\citep{park2019specAugmentAS} in all ASR runs. The models are trained using PyTorch~\citep{paszke2019pytorch} with Adam~\citep{Kingma2014AdamAM} optimizer for 120 epochs unless otherwise noted. The learning rate follows a tri-stage schedule with a maximum of 0.001. We applied gradient clipping at 10 and a weight decay parameter of 0.1. For the 960hr setting, we used a variable batch size capped at 1K utterances or 30K frames, whichever is smaller. This corresponds to about 45K update steps for 120 epochs. For the 100hr setting, we set the maximum learning rate to 0.0001 and used smaller batch size (capped at 200 utterances or 5K frames). In this case, 120 epochs corresponded to about 120K updates. For decoding, we used n-best decoding with a beam-size of 15, and evaluated the WER on the 1-best path.

\subsection{Bi-directional ALiBi Bias}\label{sec:app_alibi}
We use a symmetric variant bi-directional variant of ALiBi bias where  any query $Q_i$ and key $K_j$ with $|i-j| = N$ use the same representations \footnote{Our implementation is similar to \href{https://github.com/ofirpress/attention_with_linear_biases/issues/5}{symmetric option}.}. Furthermore, for any query $Q_i$ the bias corresponding to the flow step $x_t$ is set to 0. Similarly the bias from the flow-step $x_t$ to any other query 0. In our experiments, we find ALiBi Bias to improve convergence and extrapolation to longer sequences.
\section{Additional Experiments}

\subsection{Comparing audio model training objectives}\label{sec:app_train_obj}
While A3T is considered the regression-based speech infilling baseline, it is trained on a smaller dataset and uses a smaller model compared to \vb{}. Here we present a controlled study comparing the flow-matching and regression objectives, as well as the effectiveness of masked loss. 

We consider a reduced setup for this ablation to save the compute. All models were trained on an English audiobook dataset with 1K hours of speech using a smaller model configuration (12 layers, 1024-dimensional Transformer embedding, 2048-dimensional feed-forward layer, 8 attention heads) for 150k steps with an effective batch size of 120k frames. These models are evaluated on the cross-sentence zero-shot TTS setup (\cref{sec:exp_zs_tts}) and diverse speech sampling (\cref{sec:exp_speech_samp}). 

Results in \cref{tab:ablation_loss} show that while regression audio models produce comparable WER, the audio similarity and diversity are significantly worse. Subjective listening also reveals that the audio quality and audio similarity are much worse. On the other hand, masked loss improves audio similarity and diversity while having little impact on intelligibility. 

\begin{table}[ht]
\caption{Comparison of flow-matching and regression models, trained with loss computed on all frames or only masked frames. Results of the proposed objective is boldfaced.}
\label{tab:ablation_loss}
\centering
\begin{tabular}{ll|cc|cc}
\toprule
\multirow{2}{*}{\textbf{Method}} & \multirow{2}{*}{\textbf{Loss}} 
    & \multicolumn{2}{c|}{\textbf{Zero-Shot TTS (\textit{cross-sentence)}}}
    & \multicolumn{2}{c}{\textbf{Diverse sampling}} \\
{} & {} & \textbf{WER} & \textbf{SIM-r} & \textbf{WER} & \textbf{FSD} \\
\midrule
\colorours
Flow Matching  &  Masked & 2.1 & 0.597 & 3.1 & 242.5 \\
Flow Matching  &  All  & 2.0 & 0.528 & 3.1 & 243.1 \\
Regression  &  Masked & 2.0 & 0.520 & 2.9 & 278.8 \\
Regression  &  All  & 2.0 & 0.512 & 2.9 & 282.8 \\
\bottomrule
\end{tabular}
\end{table}

\subsection{Effectiveness on data scaling}
We create four subsets of the 60K hour English data (0.1\%, 1\%, 10\%, 100\% in duration). In particular, the $x\%$ subset would contain roughly $x\%$ of the speakers from the original set. We train one model on each subset with a reduced setup described in \cref{sec:app_train_obj} and evaluate them on zero-shot TTS (cross-sentence) and diverse sampling. Results show that scaling data constantly improves the zero-shot TTS performance (WER and SIM-r) as well as WER on diverse sampling. For FSD it shows regression when scaling from 6K hour to 60K hour, but this could result from the the reference distribution is computed from the 1K hour English audiobook data that has less diverse samples.

\begin{table}[h]
    \caption{Experiments on the effect of scaling training data.}\label{tab:scale_law}
    \centering
        \begin{tabular}{r|cc|cc}
        \toprule
        \multirow{2}{*}{\textbf{Train data (hr)}} 
            & \multicolumn{2}{c|}{\textbf{Zero-Shot TTS}}
            & \multicolumn{2}{c}{\textbf{Diverse sampling}} \\
        {} & \textbf{WER} & \textbf{SIM-r} & \textbf{WER} & \textbf{FSD} \\
        \midrule
        60      & 2.30 & 0.151 & 3.48 & 280.48 \\
        600     & 2.11 & 0.417 & 3.19 & 205.39 \\
        6,000   & 2.08 & 0.573 & 2.96 & 195.52 \\
        60,000  & 2.05 & 0.645 & 2.95 & 214.38 \\
        \bottomrule
        \end{tabular}
    
\end{table}

\subsection{Complete results on comparing generative modeling approaches}
\cref{tab:ablation_method} presents the full results of \cref{sec:exp_gen_approach} on all combinations of training steps, inference steps with results on both monolingual zero-shot TTS (\cref{sec:exp_zs_tts}) and diverse speech sampling (\cref{sec:exp_speech_samp}) for the ablation study presented in \cref{sec:exp_gen_approach}. 
In all settings Flow Matching with OT paths performs strictly better than both of the other approaches.  

\begin{table}[ht]
\caption{Comparison of FM w/OT vs. FM w/Diffusion vs. SM.}
\label{tab:ablation_method}
\centering

\begin{tabular}{lllrrrrr}
\toprule
\multirow{2}{*}{\textbf{Method}} & \multirow{2}{*}{\textbf{Train Steps}} & \multirow{2}{*}{\textbf{NFE}} & 
\multicolumn{3}{c}{\textbf{ZS-TTS (cross-sentence)}} & \multicolumn{2}{c}{\textbf{Diverse sampling}} \\
& & & \textbf{WER} & \textbf{SIM-o} & \textbf{SIM-r} & \textbf{WER} & \textbf{FSD} \\
\midrule

\multirow{12}{*}{FM w/ OT} 
         & \multirow{4}{*}{50000} 
                  & 4  &        2.7 &       0.303 &      0.362 &    4.8 &  276.499 \\
         &        & 8  &        2.5 &       0.353 &      0.412 &    4.8 &  235.958 \\
         &        & 16 &        2.4 &       0.366 &      0.425 &    4.7 &  227.485 \\
         &        & 32 &        2.5 &       0.364 &      0.424 &    4.7 &  225.931 \\
\cmidrule{2-8}
         & \multirow{4}{*}{100000} 
                  & 4   &        2.5 &       0.347 &      0.404 &    4.3 &  258.358 \\
         &        & 8   &        2.2 &       0.411 &      0.468 &    4.2 &  216.512 \\
         &        & 16  &        2.3 &       0.429 &      0.483 &    4.3 &  206.538 \\
         &        & 32  &        2.2 &       0.431 &      0.487 &    4.2 &  203.792 \\
\cmidrule{2-8}
         & \multirow{4}{*}{150000} 
                  & 4   &        2.4 &       0.356 &      0.410 &    4.0 &  249.712 \\
         &        & 8   &        2.2 &       0.430 &      0.481 &    4.0 &  208.511 \\
         &        & 16  &        2.2 &       0.453 &      0.503 &    4.0 &  198.040 \\
         &        & 32  &        2.1 &       0.458 &      0.508 &    3.9 &  195.304 \\
\midrule
\multirow{12}{*}{FM w/ diff} 
         & \multirow{4}{*}{50000} 
                  & 4   &       99.9 &       0.050 &      0.050 &   99.8 & 3478.910 \\
         &        & 8   &       99.9 &       0.047 &      0.047 &   99.9 & 4704.237 \\
         &        & 16  &       98.8 &       0.052 &      0.048 &   96.5 & 5336.591 \\
         &        & 32  &       76.0 &       0.060 &      0.066 &   49.5 & 2485.400 \\
         
\cmidrule{2-8}
         & \multirow{4}{*}{100000} 
                  & 4   &       98.9 &       0.048 &      0.048 &   96.6 & 4486.401 \\
         &        & 8   &       14.6 &       0.104 &      0.137 &   12.0 &  669.564 \\
         &        & 16  &        4.0 &       0.210 &      0.262 &    7.0 &  381.891 \\
         &        & 32  &        3.1 &       0.285 &      0.344 &    6.3 &  294.777 \\
\cmidrule{2-8}
         & \multirow{4}{*}{150000} 
                  & 4   &       11.5 &       0.132 &      0.171 &   11.4 &  692.560 \\
         &        & 8   &        3.0 &       0.305 &      0.359 &    5.6 &  334.237 \\
         &        & 16  &        2.7 &       0.391 &      0.447 &    5.4 &  244.067 \\
         &        & 32  &        2.6 &       0.423 &      0.478 &    5.2 &  224.963 \\
\midrule
\multirow{12}{*}{SM w/ diff} 
        & \multirow{4}{*}{50000} 
                  & 4   &       99.6 &       0.050 &      0.048 &   99.7 & 2816.083 \\
         &        & 8   &       99.3 &       0.051 &      0.048 &   99.6 & 3079.040 \\
         &        & 16  &       97.5 &       0.052 &      0.050 &   98.4 & 3710.340 \\
         &        & 32  &       73.3 &       0.057 &      0.062 &   86.2 & 3011.030 \\
\cmidrule{2-8}
         & \multirow{4}{*}{100000} 
                  & 4   &       99.4 &       0.050 &      0.050 &   99.3 & 3474.579 \\
         &        & 8   &       97.2 &       0.049 &      0.048 &   97.9 & 3600.423 \\
         &        & 16  &       53.9 &       0.064 &      0.071 &   69.6 & 2060.892 \\
         &        & 32  &       17.4 &       0.150 &      0.176 &   34.4 & 1071.579 \\
\cmidrule{2-8}
         & \multirow{4}{*}{150000} 
                  & 4   &       94.5 &       0.055 &      0.054 &   79.4 & 2953.417 \\
         &        & 8   &       42.3 &       0.070 &      0.076 &   27.5 & 1071.010 \\
         &        & 16  &       11.5 &       0.191 &      0.218 &   12.8 &  698.411 \\
         &        & 32  &        5.1 &       0.309 &      0.349 &    8.8 &  519.468 \\
\bottomrule
\end{tabular}

\end{table}

\subsection{Transient noise removal in more conditions}\label{sec:app_infill}
We expand the experiments in \cref{sec:exp_noise_removal} by comparing the models on two noise levels (low noise: 10dB and high noise: -10dB), three overlapping ratios (30\%, 50\%, 70\%), and also two types of noise (speech noise and non-speech noise).

Results are presented in \cref{tab:ablation_noise_removal}. \vb{} consistently produces the most intelligible audio at all conditions (indicating the percentage of speech to infill). In terms of audio similarity, \vb{} is constantly better in the high noise condition with gains ranging from 0.265 to 0.324 compared to Demucs, and is on par with Demucs in low noise condition. 

\begin{table}[ht]
\caption{Results of transient noise removal with varying overlapping percentage and noise level. ``sp'' means added noise is speech, and ``non-sp'' means non-speech.}
\label{tab:ablation_noise_removal}
\centering
\begin{tabular}{l|cccc|cccc}
\toprule
{} & \multicolumn{2}{c}{\textbf{WER}} & \multicolumn{2}{c|}{\textbf{SIM-o}} & \multicolumn{2}{c}{\textbf{WER}} & \multicolumn{2}{c}{\textbf{SIM-o}} \\
{} & sp & non-sp & sp & non-sp & sp & non-sp & sp & non-sp \\
\midrule\midrule
{}  & \multicolumn{4}{l|}{\textit{SNR=-10dB, overlap=30\%}} 
    & \multicolumn{4}{l}{\textit{SNR=10dB, overlap=30\%}} \\
Noisy speech & 26.7 & 24.9 & 0.202 & 0.238 & 3.7 & 3.1 & 0.605 & 0.603 \\
Demucs & 20.5 & 19.7 & 0.247 & 0.247 & 3.2 & 2.8 & 0.570 & 0.567 \\
A3T & \multicolumn{2}{c}{7.5} & \multicolumn{2}{c|}{0.058} 
    & \multicolumn{4}{c}{\textit{same as left}} \\
\colorours
VB-En ($\alpha=0.7$) & \multicolumn{2}{c}{2.2} & \multicolumn{2}{c|}{0.566} 
            & \multicolumn{4}{c}{\textit{same as left}} \\

\midrule\midrule
{}  & \multicolumn{4}{l|}{\textit{SNR=-10dB, overlap=50\%}} 
    & \multicolumn{4}{l}{\textit{SNR=10dB, overlap=50\%}} \\
Noisy speech & 43.6 & 40.8 & 0.256 & 0.292 & 4.5 & 3.8 & 0.649 & 0.649 \\
Demucs & 34.3 & 32.5 & 0.291 & 0.288 & 3.8 & 3.3 & 0.616 & 0.613 \\
A3T & \multicolumn{2}{c}{11.5} & \multicolumn{2}{c|}{0.064} 
    & \multicolumn{4}{c}{\textit{same as left}} \\
\colorours
VB-En ($\alpha=0.7$) & \multicolumn{2}{c}{2.0} & \multicolumn{2}{c|}{0.612} 
            & \multicolumn{4}{c}{\textit{same as left}} \\

\midrule\midrule
{}  & \multicolumn{4}{l|}{\textit{SNR=-10dB, overlap=70\%}} 
    & \multicolumn{4}{l}{\textit{SNR=10dB, overlap=70\%}} \\
Noisy speech & 60.0 & 56.0 & 0.260 & 0.303 & 6.3 & 4.6 & 0.595 & 0.592 \\
Demucs & 49.5 & 45.4 & 0.293 & 0.294 & 4.6 & 3.8 & 0.572 & 0.564 \\
A3T & \multicolumn{2}{c}{16.6} & \multicolumn{2}{c|}{0.063} 
    & \multicolumn{4}{c}{\textit{same as left}} \\
\colorours
VB-En ($\alpha=0.7$) & \multicolumn{2}{c}{2.0} & \multicolumn{2}{c|}{0.559} 
            & \multicolumn{4}{c}{\textit{same as left}} \\
\bottomrule
\end{tabular}
\end{table}

\subsection{Choice of audio model output features}
The performance of our model is upper bounded by how well the chosen acoustic features can be reconstructed to waveform. The reconstruction performance is determined jointly by the encoding process, as in how much information is lost when encoding waveform into the features, and the decoding process, as in how well the vocoder can translate the encoded information into waveform.

To motivate the choice of the acoustic feature and the vocoder, we compare four combinations: the first one is Mel spectrogram + HiFi-GAN which is what this paper adopts. The second is Mel spectrogram + Parallel WaveGAN~\citep{yamamoto2020parallel} that is used by A3T~\citep{Bai2022A3TAA}. The third one is Encodec post-quantization dense feature + Encodec decoder, which is analogous to VALL-E's setup. The last one is also Encodec but with pre-quantization dense feature, which we include to study how much information is lost during quantization.

We also note that Mel spectrogram features are 80 dimensional encoded at 100Hz, which is 8K dimensions per second, while Encodec features are 128 dimensional encoded at 75Hz, which is 9.6K dimensions per second, higher than the Mel spectrogram features.

\cref{tab:ablation_audio_feat} presents the results evaluated on the Librispeech dev-clean and dev-other splits. All three models have the same WER resynthesizing dev-clean split, but ParallelWaveGAN degrades the most on dev-other. Interestingly Encodec even produces audio of lower WER than the ground truth. 

In terms of audio similarity, besides the default audio feature extractor WavLM-TDCNN, we also include results of similarity computed with another speaker encoder ECAPA~\citep{desplanques2020ecapa}. Parallel WaveGAN is consistently the worst. However, it is unclear whether HiFi-GAN or Encodec performs better. Encodec prevails with the WavLM-TDCNN embedder and HiFi-GAN wins using ECAPA. It may require subjective MOS test to conclude which one reconstructs the audio better, and we leave exploration of modeling Encodec dense features for future study.

\begin{table}[h]
\caption{Comparison of different audio features and vocoders on audio reconstruction. Librispeech dev-clean (d-c) and dev-other (d-o) are used for evaluation. WER and audio similarity computed with WavLM-TDCNN and ECAPA are reported.}
\label{tab:ablation_audio_feat}
\centering
\resizebox{\linewidth}{!}{
\begin{tabular}{l|cccccc}
    \toprule
    \multirow{2}{*}{\textbf{Audio feature / Vocoder}}
    & \multicolumn{2}{c}{\textbf{WER}}
    & \multicolumn{2}{c}{\textbf{SIM-o (WavLM)}}
    & \multicolumn{2}{c}{\textbf{SIM-o (ECAPA)}} \\
    {} & d-c & d-o & d-c & d-o & d-c & d-o \\
    \midrule

Ground truth & 2.1 & 4.7 & 1.000 & 1.000 & 1.000 & 1.000 \\
\midrule
\colorours
Mel spectrogram / HiFi-GAN & 2.1 & 4.7 & 0.915 & 0.909 & 0.766 & 0.762 \\
Mel spectrogram / Parallel WaveGAN & 2.1 & 5.2 & 0.868 & 0.847 & 0.721 & 0.711 \\
Encodec post-quantized feature / Encodec decoder & 2.1 & 4.5 & 0.943 & 0.944 & 0.724 & 0.722 \\
Encodec pre-quantized feature / Encodec decoder & 2.1 & 4.4 & 0.943 & 0.944 & 0.724 & 0.722 \\

    \bottomrule
\end{tabular}
}
\end{table}

\section{Additional Details and Studies on Metrics}

\subsection{Measuring speech diversity and quality with FSD}\label{sec:app_fsd_ablation}

\paragraph{Diversity}
We first validate if FSD reflects the diversity for a set of speech samples and study its sensitivity to sample size. To achieve that, we design controlled experiments to compute FSD on sets of samples with varying diversity and sample sizes. Specifically, we create two partitions from 1K hours of English speech, where each partition has the same set of speakers and the same number of utterances for each speaker. The first partition is considered the reference set.

To test the sensitivity to sample size, we use the second partition to create subsets by sampling $r\%$ of utterances from each speaker in that partition. This sampling method is denoted as ``utt''. We computed that on average, each speaker contributed approximately $2.33$ sessions, with each session containing around $52.45$ utterances. Therefore, the subsets created using the sampling method are expected to have similar audio style distributions to the reference set and the FSD is expected to stay low regardless of the subset size. We consider $r \in \{1, 5, 10, 25, 50, 100\}$.

To test the correlation with diversity, we again use the second partition to create subsets by sampling $r\%$ of speakers and including all the utterances in the partition from those speakers. This sampling method is denoted as ``spk'' where a smaller $r$ leads to a subset with fewer speakers and hence lower diversity. Therefore the FSD is expected to increase as $r$ decreases. The same set of values for $r$ is considered. For the same $r$, the ``utt'' subset should always have a lower FSD than the ``spk'' subset.

We compare three different features for computing the FSD score. The first is the supervised WavLM-TDCNN feature used for computing audio similarity (SIM-r and SIM-o). The second is the self-supervised wav2vec 2.0 \textsc{Base}~\citep{baevski2020wav2vec} feature reduced to 128 dimensions using principle component analysis (PCA). The last one is the supervised audio event classification model feature that is used to compute FAD~\citep{Kilgour2019FrchetAD} for non-speech audio generation.

\begin{figure}[htp]
    \centering
    \includegraphics[width=\linewidth]{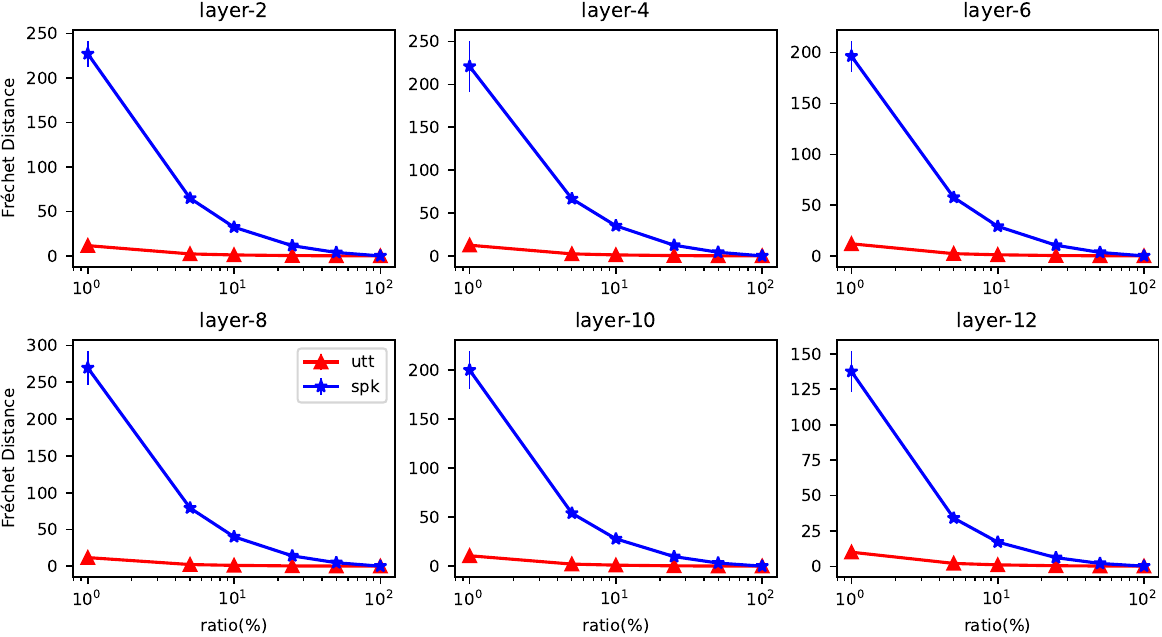}
    \caption{FSD based on different layers of wav2vec 2.0 \textsc{Base}. utt: utterance-based sampling, spk: speaker-based sampling. Vertical bars denote standard deviation.}
    \label{fig:metric-fsd-layer}
\end{figure}

Figure~\ref{fig:metric-fsd-layer} first compares using different layers of wav2vec 2.0 features. All of them yield similar desirable results where ``utt'' stays low and ``spk'' increases drastically when the sample size reduces and speaker diversity decreases. We then decide to use the middle layer (layer 6) as the default feature for FSD computation.

\begin{figure}[htp]
    \centering
    \includegraphics[width=\linewidth]{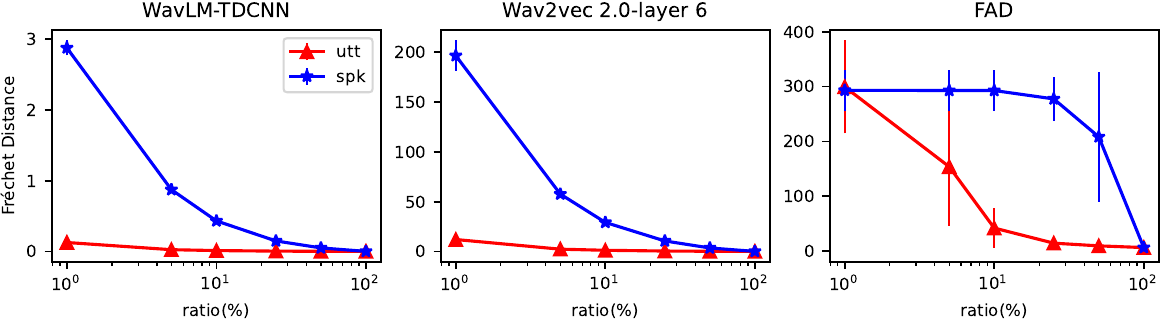}
    \caption{FSD with different sample size using supervised WavLM-TDCNN, self-supervised wav2vec 2.0, and supervised audio event classifier features. utt: utterance-based sampling, spk: speaker-based sampling. Vertical bars denote standard deviation.}
    \label{fig:metric-fsd-feature}
\end{figure}

Figure~\ref{fig:metric-fsd-feature} further compares wav2vec 2.0-layer 6 with the two other features. WavLM-TDCNN and wav2vec 2.0-layer 6 present similar trends and both have low variance. Both of them are suitable for measuring diversity, and we decide to use wav2vec 2.0 features as it is self-supervised and would be able to capture more holistic information of speech such as prosody and emotion. 

In contrast, FAD score~\citep{Kilgour2019FrchetAD} is not appropriate for measuring speech diversity. The score does not increase much between $r=25\%$ and $r=1\%$ for ``spk'' sampling method, showing that the score does not reflect the decreasing speaker diversity. On the other hand, ``utt'' sampling method observes huge FAD score increase when reducing the sample size from $r=25\%$ to $r=1\%$ where the diversity does not change much as the number of speakers remains the same. Moreover, at $r=1\%$ both sampling methods result in similar FAD score while the two subsets exhibit very different levels of diversity. We hypothesize that this is because FAD score is computed based on features extracted from an audio even classifier trained on AudioSet, which learns to distinguish between events like lawn mower, car engine, and human speech, but does not learn to capture the variation within speech, such as different voices.

\paragraph{Quality}
In addition to measuring diversity, Fr\'echet distance is a commonly used metric for assessing quality in image generation~\citep{Ho2020DenoisingDP}. To show its applicability for speech generation, we evaluate the FSD score of speech utterances with varying levels of quality. 
The reference set samples are 1K hours of English training data, and the hypothesis set is the Librispeech test-clean split with noise added.
We added Gaussian noise at different SNRs, ranging from 0 to 50 dB. Lower SNR values correspond to lower quality. We use the default speech feature extractor (i.e., wav2vec 2.0, layer-6) throughout the experiments. 

\begin{figure}[htp]
    \centering
    \includegraphics[width=0.8\linewidth]{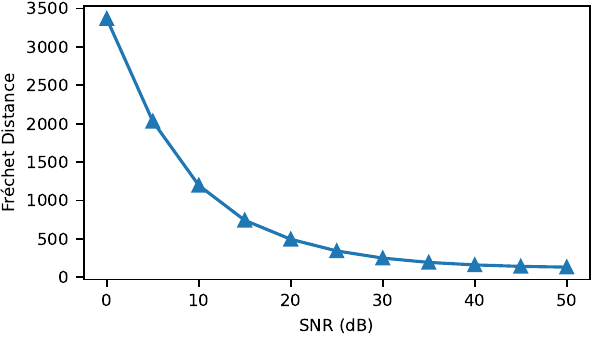}
    \caption{FSD under different noisy levels. Feature: Wav2vec 2.0 layer-6 feature. Noise is added upon model output from \vb{} under unconditional setting.}
    \label{fig:metric-fsd-snr}
\end{figure}

Our results, summarized in Figure~\ref{fig:metric-fsd-snr}, show that a subset with a lower SNR has a higher FSD score. Therefore, a lower FSD score indicates higher acoustic quality for the set of test samples when diversity is fixed.
 
\subsection{Standalone metrics for duration models}\label{sec:app_dur_metrics}
As mentioned in the main text, we can utilize end-to-end metrics of WER, SIM, and FSD to evaluate duration models, but also consider metrics specifically for duration. 

First, we consider two metrics aimed at the quality of duration predictions, here denoted $\hat{l}(\lctx, y)$.  For a regression model, we use $\hat{l}(\lctx, y) = g(\lctx, y; \theta)$.  For a flow matching model, we set $\hat{l}$ as the mean over $20$ samples, ensuring a fairer comparison.  

\paragraph{Duration correctness (MS-MAE)} Our first metric, multi-sample mean-absolute error (MS-MAE), is the masked absolute error per-utterance divided by the average number of masked phonemes per-utterance
\begin{align}
\frac{\mathbb{E}_{m, l, y} || m \odot \left( l - \hat{l}(\lctx, y) \right) ||_1} {\mathbb{E}_{m, l, y} ||m||_1}
\end{align}

\paragraph{Speaking rate correlation (MS-Corr)} Our next metric, multi-sample correlation (MS-Corr), computes the average masked predicted duration and unmasked duration context per utterance, and computes their correlation across utterances. Comparing MS-Corr with the same correlation computed from the ground truth, we observe to what extent predicted durations capture appropriate correlations with the context.

\paragraph{Duration diversity and quality (FDD)} Additionally, we evaluate the quality and diversity of duration samples at the distribution level, similar to our audio evaluation of diversity and quality via FSD.  We produce one sample per utterance from a duration model and collect all sampled phoneme durations, possibly many per-utterance, into an empirical distribution.  We compare means and variances of this sampled distribution versus the means and variances of the training distribution, labeled $\mu$, $s$, and $\mu'$, $s'$ respectively. We define the Fr\'echet duration distance (FDD) as the Fr\'echet distance between the distributions 
\begin{align}
(\mu - \mu')^{2} + s + s' - 2 \sqrt{s s'},
\end{align}
treated as though they were Gaussians.  FDD depends on the sampled durations accurately reflecting the training distribution of real durations.  As for FSD, this metric is specific to unconditional text-to-speech generation.

\subsection{Duration model evaluation with standalone metrics}\label{sec:app_dur_metrics_eval}
We evaluate three duration model variants.  The first and second utilizes flow matching and regression, trained using masked conditional flow matching and regression respectively as described in Section~\ref{subsec:model_and_training}.  The third is a regression model that ignores duration context $\lctx$ and only uses phonetic transcript $y$, referred to as unconditional regression below. This is the duration model used in FastSpeech2~\citep{Ren2020FastSpeech2F}, A3T~\citep{Bai2022A3TAA} and many other non-autoregressive speech synthesis models.

We evaluate our three duration model variants on the Librispeech test-other split on two tasks.  The first is unconditional TTS where we generate all durations from given phonemes (i.e. $\lctx$ is entirely masked).  The second task is infilling the second half of each utterance's durations, where $\lctx$ are durations from the unmasked half of the utterance. This second infilling task distinguishes between the two regression model variants, since the unconditional regression ignores $\lctx$, and hence predicts identical durations for the tasks.  Duration metrics are computed for TTS and infilling in Table~\ref{tab:duration_eval_tts} and~\ref{tab:duration_eval_infilling}. The prefix Phn or Sil indicates the associated metric was either computed across all non-silence or all silence phonemes.  Start and end silences were not trimmed for these duration metric evaluations.

Starting with prediction quality metrics (MS-MAE and MS-Corr), the duration-conditional regression performs slightly better on MS-MAE overall than the other models. Larger differences are seen on Phn-MS-Corr where the unconditional regression has a correlation substantively below the other models (Phn-MS-Corr of ground truth is $0.47$), indicating conditioning on duration context $\lctx$ is beneficial.  Flow-matching shows the largest distinction versus regression on the distributional comparison captured by FDD.  The regression models have generally larger FDD because they underestimate the standard deviation in phoneme and silence durations, and hence produce samples with less duration diversity and more regular duration lengths. 

\begin{table}[ht]
    \caption{English TTS duration metrics on LS test-other.}
    \label{tab:duration_eval_tts}
    \centering
    \resizebox{\linewidth}{!}{
    \begin{tabular}{l|cccc}
    \toprule
    \textbf{Duration Model} & \textbf{Phn-MS-MAE} & \textbf{Phn-FDD} & \textbf{Sil-MS-MAE} & \textbf{Sil-FDD} \\
    \midrule
    Unconditional Regression & 2.53 & 0.72 & 5.32 & 2.39 \\
    Duration-conditional Regression & 2.52 & 0.76 & 5.10 & 8.40 \\
    Duration-conditional Flow Matching & 2.63 & 0.61 & 5.18 & 2.48 \\
    \bottomrule
    \end{tabular}
    }
\end{table}

\begin{table}[ht]
    \caption{English second-half infilling duration metrics on LS test-other.}
    \label{tab:duration_eval_infilling}
    \centering
    \begin{tabular}{l|ccc}
    \toprule
    \textbf{Duration Model} & \textbf{Phn-MS-MAE} & \textbf{Phn-MS-Corr} & \textbf{Sil-MS-MAE} \\
    \midrule
    Unconditional Regression & 2.57 & 0.26 & 5.44 \\
    Duration-conditional Regression & 2.45 & 0.35 & 5.20 \\
    Duration-conditional Flow Matching & 2.52 & 0.41 & 5.32 \\
    \bottomrule
    \end{tabular}
\end{table}

\subsection{Duration model evaluation with end-to-end metrics}\label{sec:app_dur_e2e_eval}
We now present end-to-end metrics for our three duration variants for  zero-shot TTS cross-sentence and continuation, as well as diverse speech generation, corresponding to Sections~\ref{sec:exp_zs_tts} and~\ref{sec:exp_speech_samp}.  Zero-shot TTS cross-sentence and continuation results are shown in Table~\ref{tab:tts_unfiltered_duration} and diverse speech generation results in Table~\ref{tab:samp_duration}. These results are not comparable with the main text as they utilize the flow-matching model described in Appendix~\ref{sec:app_train_obj}, denoted as VB-En-1K.

Overall, FSD and SIM are similar across duration variants.  On the other hand, WER is sensitive to the choice of duration model, where the duration-conditional regression achieves a substantially lower WER.  Subjective listening from the duration-conditional regression and flow-matching confirms that the regression model is producing more regular patterns of speech, that may be easier for ASR to recognize, while sacrificing some duration diversity. 

\begin{table}[ht]
    \caption{Diverse speech generation from LS test-other text.}
    \label{tab:samp_duration}
    \centering
    \begin{tabular}{l|cc}
    \toprule
    \textbf{Duration Model with VB-En-1K} & \textbf{WER} & \textbf{FSD (LS-train)} \\
    \midrule
    Unconditional Regression & 3.8 & 148.7 \\
    Duration-conditional Regression & 3.7 & 148.1 \\
    Duration-conditional Flow Matching & 5.4 & 155.1 \\
    \bottomrule
    \end{tabular}
\end{table}

\begin{table}[ht]
    \caption{English zero-shot TTS results on filtered LS test-clean.}
    \label{tab:tts_unfiltered_duration}
    \centering
    \begin{tabular}{l|ccc}
    \toprule
    \textbf{Duration Model with VB-En-1K} & \textbf{WER} & \textbf{SIM-o} & \textbf{SIM-r} \\
    \midrule
    \multicolumn{4}{l}{\textit{cross-sentence}} \\
    Unconditional Regression & 3.0 & 0.538 & 0.584 \\
    Duration-conditional Regression & 2.7 & 0.545 & 0.591 \\
    Duration-conditional Flow Matching & 3.4 & 0.528 & 0.578 \\
    \midrule
    \multicolumn{4}{l}{\textit{continuation}} \\
    Unconditional Regression & 2.5 & 0.485 & 0.524 \\
    Duration-conditional Regression & 2.2 & 0.491 & 0.533 \\
    Duration-conditional Flow Matching & 2.7 & 0.481 & 0.525 \\
    \bottomrule
    \end{tabular}
\end{table}

\subsection{MOS instructions}\label{sec:app_mos_instruct}
\cref{tab:qmos} shows the instruction presented  to the raters for quality mean opinion score study. \cref{tab:smos} shows the instruction presented  to the raters for similarity mean opinion score study.

\begin{table}[ht]
    \centering
    \caption{Quality mean opinion score (QMOS) instruction.}
    \label{tab:qmos}
\begin{tabular}{p{12cm}}
\toprule\toprule
\textbf{Introduction} \\
Your task is to evaluate the subjective quality and intelligibility of the speech from short (2-8 second) audio files. Each HIT can be completed in roughly around 120 seconds. \\
\\
\textbf{Task Instructions} \\
In this task you will hear samples of speech recordings. The purpose of this test is to evaluate the quality and intelligibility of each file in terms of its overall sound quality and the amount of mumbling and unclear phrases in the recording. \\
\\
Please keep in mind that speech samples can be distorted and noisy, however these are only specific examples. \\
\\
Please use a headset for listening and adjust your volume level to your comfort during this training, and do not change later during the experiment. \\
\\
You should give a score according to the following scale, known as the MOS (mean opinion score) scales:\\
\\
\textbf{Score (Quality and Intelligibility of the speech) }\\
5	(Excellent) \\
4	(Good) \\
3	(Fair) \\
2	(Poor) \\
1	(Bad) \\
\bottomrule\bottomrule
\end{tabular}
\end{table}

\begin{table}[ht]
    \centering
    \caption{Similarity mean opinion score (SMOS) instruction.}
    \label{tab:smos}
\begin{tabular}{p{12cm}}
\toprule\toprule
\textbf{Task Name} \\
Rate the similarity of the synthesized speech samples to a given prompt. \\
\\
\textbf{Task Instructions} \\
Your task is to evaluate the similarity of the synthesized speech samples to the given speech prompt. You should focus on the similarity of the speaker, speaking style, acoustic conditions, background noise, etc. You should rank the recordings on the scale between 1-5, where 5 is the best quality and 1 is the worst. \\
\\
In other words, please rank the recordings according to their acoustic similarity to the given prompt, meaning as if they were recorded in the same place by the same speaker speaking in similar styles. This task typically requires approximately 120 seconds to complete.\\
\\
Please use a headset for listening and adjust your volume level to your comfort during this training, and do not change later during the experiment. \\
\bottomrule\bottomrule
\end{tabular}

\end{table}

\section{Detailed Configurations for Acoustic and Duration model training}

\begin{table}[H]
\centering
\caption{Detailed configurations for the audio models used in our experiments.}
\label{tab:configurations}
\begin{tabular}{l|l|l}
\toprule
 & {VB-En} & {VB-Multi} \\
\midrule
\multicolumn{3}{c}{Model Parameters}\\
\midrule
Model Dimension & 1024 & 1024 \\
Number of Heads & 16 & 16 \\
Number of Layers & 24 & 24 \\
Feedforward Dimension & 4096 & 4096 \\
Attention Dropout & 0.0 & 0.0 \\
Activation Dropout & 0.1 & 0.1 \\
ConvPos Width & 31 & 31 \\
ConvPos Groups & 16 & 16 \\
ConvPos Depth & 2 & 2 \\
Skip Connections & true & true \\
Alibi Bias & true & true \\
\midrule
\multicolumn{3}{c}{Training Parameters}\\
\midrule
Number of Iterations & 500000 & 750000 \\
Number of GPUs & 32 & 32 \\
Learning Rate (LR) & 0.0001 & 0.0001 \\
Gradient Clipping Value & 0.2 & 0.2 \\
LR Scheduler Warmup Steps & 5000 & 5000 \\
Loss Masking & true & true \\
\midrule
\multicolumn{3}{c}{Data Parameters}\\
\midrule
Tokens per Batch & 7500 & 7500 \\
Conditional Dropout & 0.2 & 0.2 \\
Position Dependent Phones & true & true \\
Phoneme Mask Percent & 0.0, 0.0 & 0.0, 0.0 \\
Spectrogram Mask Percent & 0.7, 1.0 & 0.7, 1.0 \\
Spectrogram Drop Percentage & 0.3 & 0.3 \\
Chunk Length & 1600 & 1600 \\
Transform Type & normalize & normalize \\
Mean & -5.884 & -5.884 \\
Standard Deviation & 2.261 & 2.261 \\
Upsampling $\beta$ & - & 0.25 \\
\bottomrule
\end{tabular}
\end{table}

\begin{table}
\centering
\caption{Detailed configurations for conditional flow matching based duration models used in our experiments. For regression based model, we use the same configurations but use regression loss.}
\label{tab:configurations}
\begin{tabular}{l|l|l}
\toprule
 & {VB-En} & {VB-Multi} \\
\midrule
\multicolumn{3}{c}{Model Parameters}\\
\midrule
Model Dimension & 512 & 768 \\
Number of Layers & 8 & 10 \\
Feedforward Dimension & 2048 & 2048 \\
Attention Dropout & 0.1 & 0.1 \\
Activation Dropout & 0.1 & 0.1 \\
ConvPos Width & 15 & 15 \\
ConvPos Groups & 16 & 16 \\
ConvPos Depth & 2 & 2 \\
Skip Connections & true & true \\
Alibi Bias & true & true \\
\midrule
\multicolumn{3}{c}{Training Parameters}\\
\midrule
Number of Iterations & 600000 & 600000 \\
Number of GPUs & 4 & 4 \\
Learning Rate (LR) & 0.0001 & 0.0001 \\
LR Scheduler Warmup Steps & 5000 & 5000 \\
Loss Masking & true & true \\
\midrule
\multicolumn{3}{c}{Data Parameters}\\
\midrule
Conditional Dropout & 0.2 & 0.2 \\
Upsampling $\beta$ & - & 0.5 \\
Tokens per Batch & 15000 & 15000 \\
Duration Drop Percentage & 0.2 & 0.2 \\
Duration Mask Percent & 0.1, 1.0 & 0.1, 1.0 \\
Position Dependent Phones & true & true \\
Transform Type & log & log \\
\bottomrule
\end{tabular}
\end{table}

\end{document}